\documentclass[aps,prc,groupedaddress,onecolumn,floatfix]{revtex4}
\usepackage{amsmath}
\usepackage{amssymb} 
\usepackage{amsfonts}
\usepackage{amscd}
\usepackage{bm}
\usepackage{graphicx}
\usepackage{color}
\usepackage{hyperref}
\hypersetup{colorlinks=true, citecolor=blue, urlcolor=blue}
\usepackage{empheq}

\def\beq{\begin{equation}}
\def\eeq{\end{equation}}

\begin{document}

\title{Light-front puzzles} 

\author{W.~N.~Polyzou}
\email{polyzou@uiowa.edu}
\thanks{This work supported by the U.S. Department of Energy,
  Office of Science, Grant \#DE-SC0016457}
\affiliation{Department of Physics and Astronomy\\ The University of
Iowa\\ Iowa City, IA 52242, USA}

\date{\today}

\begin{abstract}

Light-front formulations of quantum field theories have many advantages
for computing electroweak matrix elements of strongly interacting
systems and other quantities that are used to study hadronic
structure.  The theory can be formulated in Hamiltonian form so
non-perturbative calculations of the strongly interacting initial and
final states are in principle reduced to linear algebra. These states
are needed for calculating parton distribution functions and other
types of distribution amplitudes that are used to understand the
structure of hadrons.  Light-front boosts are kinematic
transformations so the strongly interacting states can be computed in
any frame.  This is useful for computing current matrix elements
involving electroweak probes where the initial and final hadronic
states are in different frames related by the momentum transferred by
the probe.  Finally in many calculations the vacuum is trivial so the
calculations can be formulated in Fock space.

The advantages of light front-field theory would not be interesting if
the light-front formulation was not equivalent to the covariant or canonical
formulations of quantum field theory.  Many of the distinguishing properties of
light-front quantum field theory are difficult to reconcile with
canonical or covariant formulations of quantum field theory.

This paper discusses the resolution of some of the apparent
inconsistencies in canonical, covariant and light-front formulations
of quantum field theory.  The puzzles that will be discussed are (1)
the problem of inequivalent representations (2) the problem of the
trivial vacuum (3) the problem of ill-posed initial value problems (4)
the problem of rotational covariance (5) the problem of zero modes and
(6) the problem of spontaneously broken symmetries.

\end{abstract}

\maketitle 

\section{Introduction}\label{sec1}

Light-front formulations of quantum field theory
\cite{Chang1069}\cite{Soper}\cite{Chang:1972xt}\cite{Yan:1973.2}\cite{Yan:1973.3}\cite{Yan:1973.4}
have a number of advantages for studying the structure of hadrons.
Among the most important advantages are (1) it has a Hamiltonian formulation
\cite{VARY201064}  so hadronic state vectors can in principle be found using methods of
linear algebra.
The Hamiltonian formulation
has the advantage that it is non-perturbative.  The
resulting hadronic wave-functions can be used to compute form factors,
parton distribution functions, generalized parton distribution
functions and related quantities that are important for unraveling
hadronic structure (2) there is a three-parameter group of
interaction-independent boosts that is useful for studying electroweak and
gravitational probes of hadronic systems, where the momentum
transferred by the probe puts the initial and final hadronic states in
different reference frames.  Because the light-front boosts form a
subgroup the light-front spins do not Wigner rotate under light-front
boosts.  (3) The algebra of fields restricted to the light front is
irreducible and the vacuum is trivial. This implies that calculations
can be formulated directly on the free field Fock space of the theory.

These advantages are in contrast to properties of canonical or
covariant formulations of quantum field theory.  The light-front
formulation would not be interesting if it was not equivalent to these
conventional formulations of quantum field theory, however some the
distinguishing properties of light-front field theory are difficult to
reconcile with properties of canonical or covariant formulations of
quantum field theory.  Some of these puzzles include:

\begin{itemize}

\item[$\bullet$] The problem of inequivalent representations of the canonical commutation
relations.

\item[$\bullet$] The problem of the trivial vacuum.

\item[$\bullet$] The problem of the initial value problem.
  
\item[$\bullet$] The problem of rotational covariance.
  
\item[$\bullet$] The problem of zero modes.

\item[$\bullet$] The problem spontaneously broken symmetries.

\end{itemize}
A brief description of each of these problems is given below.

1) The problem of inequivalent representations: For systems with a
finite number of degrees of freedom the Stone-Von Neumann theorem
\cite{MHStone1}\cite{JvNeumann2}\cite{JvNeumann3} implies that
canonical transformations can be realized by unitary transformations.
Unitary transformations preserve the canonical commutation relations,
$[p_i,q_j]= -i\delta_{ij}$ and
$[p_i,p_j]=[q_i,q_j] =0$.  For harmonic oscillators with different
frequencies the unitary transformation relating the momentum and
coordinate of one oscillator to the momentum and coordinate of the other
oscillator transforms the creation and annihilation operators for one
oscillator to linear combinations of creation and annihilation
operators for the other oscillator.  It also transforms the vacuum
vector, which is annihilated by the transformed annihilation operator.
Canonical field theories are canonical systems with an infinite number
of degrees of freedom.  Canonical coordinates and momenta can be
obtained by integrating the fields and generalized momenta at fixed
time with an orthonormal set of basis functions.  The Stone-Von
Neumann theorem does not hold for systems with an infinite number of
degrees of freedom.  Haag \cite{Haag:1955ev} has given a simple
example which demonstrates the breakdown of this theorem for free
field theories with different masses.  What happens is that the
generator of the unitary transformation that relates the creation and
annihilation operators and ground states for systems of a finite
number of degree of freedom maps the ground state of one system to a
vector whose norm becomes infinite in the limit of an infinite number
of degrees of freedom.  On the other hand a simple change of variables
transforms the canonical fields to light-front fields.  In the light-front
case there is a kinematic representation of the irreducible Weyl
algebra (exponential form of the canonical commutation relations), and
the Weyl algebra and ground states for free field theories with
different masses are unitarily equivalent
\cite{Schlieder:1972qr}. Thus, while the vacua of two canonical free
field theories with different masses are vectors in inequivalent Hilbert space
representations of the canonical commutation relations, in the
light-front representation the Hilbert space representations are
unitarily equivalent.  This does not seem to be consistent with the
assumption that the canonical and light-front formulations are
equivalent.

2) The problem of the triviality of the light-front vacuum: The
spectral condition $P^+\geq 0$ in light-front quantum field theory
suggests that interactions cannot change the vacuum.
This follows because the interaction must commute with
$P^+$ which means that the interaction applied to the vacuum
is an eigenstate of $P^+$ with eigenvalue 0.  This implies that
the interaction cannot change the vacuum.
This is
consistent with the observation that vacuum diagrams are suppressed by
boosting to a frame with large momentum
\cite{Weinberg:2009ca}\cite{PhysRev.158.1638}.  On the other hand it
is known that in canonical formulations of quantum field theory for
Hamiltonians that are quadratic in the generalized momenta the vacuum
determines the Hamiltonian \cite{Araki:1964}\cite{PhysRev.117.1137}.
Similarly, in algebraic formulations of quantum field theory, all that
is known about the fields is that they are local operator valued
distributions that transform covariantly.  The vacuum is a linear
functional on the algebra of smeared field operators.  The dynamics is given
by specifying the Wightman distributions of the theory which are
vacuum expectation values of products of the fields.  Different vacuum
functionals acting on the same local field algebra result in different
Wightman distributions, which result in unitarily inequivalent
theories.  One
interpretation of these observations is that the vacuum defines the
dynamics in conventional formulations of quantum field theory, which
is in contrast to the light-front case where all theories have equivalent
vacua.

3) The problem of an ill-posed initial value problem: A fixed time
hyperplane is a good initial value surface.  The fields and their time
derivatives (generalized momenta) at fixed time are an irreducible set
of operators.  They can be used to construct initial data. These
operators can be evolved in time using the Heisenberg field equations.
A light front is a hyperplane that is tangent to the light cone. It
includes points that are causally related by a light signal, so it is
not a suitable initial value surface.  While massive particles do not
move at the speed of light, there is still missing information that
must be supplied on the light front in order to get a well-posed
initial value problem.  On the other hand, because the fields
restricted to a light front are irreducible they can also be used to
represent any operator, including all of the Poincar\'e generators.
Initial data can also be represented by applying operators in the
irreducible light-front field algebra to the light-front vacuum.  The
evolution of this data normal to the light front is given by a unitary
one-parameter subgroup of the Poincar\'e group, which suggests that
the initial value problem is well posed.

4) The problem of rotational covariance.  The Poincar\'e Lie algebra
has 10 infinitesimal generators.  In light-front representations there
are 7 interaction-independent generators (kinematic generators) and
three dynamical generators.  The dynamical generators can be chosen as
the light-front Hamiltonian $P^-:= H-\mathbf{P}\cdot \hat{\mathbf{z}}$
and 2 transverse rotation $\hat{\mathbf{z}} \times \mathbf{J}$
generators.  The light-front Hamiltonian and the kinematic generators
form a closed Lie algebra.  This means that the light-front
Hamiltonian does not intrinsically have information on how to
formulate rotational covariance.  In addition the construction of the
Poincar\'e Lie algebra is not unique.  Specifically if $J^1$ and $J^2$
are a pair of transverse rotation operators that complete the
Poincar\'e Lie algebra and $W$ is a unitary operator that commutes
with $P^-$ and the kinematic generators, then $J^{1\prime}:=
WJ^1W^{\dagger}$ and $J^{2\prime}:= WJ^2W^{\dagger}$ are a different
set of transverse rotation operators that complete the Poincar\'e Lie
algebra.  On the other hand Noether's theorem on the light front
results in explicit expressions for the transverse rotation
generators.  This means that eigenstates of $P^-$ are not
intrinsically constrained by rotational covariance.  This results in a
problem in assigning spins to states that are degenerate in
light-front magnetic quantum numbers.

5) The problem zero modes: In light front-quantum field theory there
are the standard divergences associated with large momentum and new
divergences that appear as $p^+\to 0$.  These divergences are related
by both rotational covariance and space reflection symmetry.  This is
because the divergences of a quantum field theory associated with
$p^3\to -\infty$ become divergences as $p^+\to 0$ in light-front
quantum field theory.  Both the large momentum and small $p^+$
divergences need to be removed.  The remaining finite parts are
constrained by rotational covariance and space reflection symmetry.
This follows because in light-front quantum theories full rotational
invariance is equivalent to the requirement that changing the
orientation of the light front leaves all physical observables
unchanged \cite{Karmanov1}\cite{Karmanov2}
\cite{Karmanov3}\cite{Karmanov4}\cite{Karmanov5}\cite{Fuda:1994uv}
\cite{Fuda:1990}\cite{Polyzou:1999}.  Changing the orientation of the
light front implies that the $p^+\to 0$ divergence associated with one
light front is transformed a large momentum divergence in an 
equivalent theory with a different light front.  After renormalization
the finite parts must be consistent with rotational covariance.

Similar considerations apply to a reflection about the $x-y$ plane.
This is a dynamical transformation that transforms $P^+$ to $P^-$.
Space reflection relates the $p^+\to 0 $ divergences to $p^3 \to
\infty$ divergences.  Thus space reflection symmetry, like rotational
symmetry, implies that the renormalization of the $p^+\to 0$ and
$\vert \mathbf{p}\vert \to \infty$ divergences cannot be performed
independently; they are constrained by rotational covariance and space
reflection symmetry.

These considerations imply that separately renormalizing the $p^+\to
0$ and large-momentum singularities in $P^-$ is not sufficient to
restore rotational covariance and space reflection symmetry.
It is known that additional zero-mode contributions are required to
reconcile some calculations in light-front quantum field theory with
calculations based on covariant perturbation theory
\cite{Maskawa:1976} \cite{Yamawaki:1998}\cite{choi:1998} where both
rotational and space reflection symmetry are satisfied.

6) The problem of spontaneous symmetry breaking: In canonical field
theories spontaneous symmetry breaking is due to a charge operator
that couples the vacuum to a massless Goldstone boson, while in the
light-front case the spectral condition $P^+\geq 0$ implies that
the light-front charge operator leaves the vacuum invariant.

The advantages of light-front quantum, field theory and its relation
to covariant and canonical quantum field theory as well as the analysis of some of the
problems listed above have been the subject of many papers
in the literature
\cite{Chang1069}
\cite{Soper}
\cite{Leutwyler:1970wn}
\cite{Rohrlich:1971zz}
\cite{Chang:1972xt}
\cite{Yan:1973.2}
\cite{Yan:1973.3}
\cite{Yan:1973.4}
\cite{Schlieder:1972qr}
\cite{Maskawa:1976}
\cite{Nakanishi:1976yx}
\cite{Karmanov1}
\cite{Karmanov2}
\cite{Nakanishi:1977}
\cite{Leutwyler:1977vy}
\cite{Karmanov3}
\cite{Karmanov4}
\cite{Karmanov5}
\cite{Fuda:1990}
\cite{Coester:1992}
\cite{Coester:1993fg}
\cite{wilson:1994}
\cite{Fuda:1994uv}
\cite{Bylev:1996}
\cite{Brodsky:1998}
\cite{Yamawaki:1998}
\cite{choi:1998}
\cite{Tsujimaru:1997jt}
\cite{Polyzou:1999}
\cite{Lenz:2000}
\cite{Heinzl:2001}
\cite{Burkhardt:2002}
\cite{brodsky_srivastava_1},
\cite{brodsky_2002}
\cite{Heinzl:2003}
\cite{Werner:2006}
\cite{lubo2008higgs}
\cite{Bakker:2011zza}
\cite{Choi:2011xm}
\cite{Choi:2013ira}
\cite{Beane:2013oia}
\cite{sofia:2014}
\cite{Herrmann:2015dqa}
\cite{BRODSKY20151}
\cite{Hiller:2016}
\cite{Ji:2017vfu}
\cite{chueng_1}
\cite{Collins:2018aqt}
\cite{Mannheim:2020rod}
\cite{chueng_2}
\cite{polyzou:2021}
\cite{brodsky_2022}.
The purpose of this work is reconcile some of these puzzles; many of which have
been previously discussed in theses references. 

Free field theories provide examples of solvable quantum field
theories where some of these problems can be investigated.  Real
applications require interacting field theories.  These are best
understood using perturbation theory.  Axiomatic approaches can
provide some insight to some of these problems that are not tied to
perturbation theory, but as a practical matter there are unresolved
questions to how to renormalize light-front theories and restore
rotational covariance and reflection symmetry outside of perturbation
theory.

A brief summary of the observations and resolutions of some of the
puzzles discussed above is given below.  They will be discussed in more
detail in what follows.

\begin{itemize}
  
\item[1.] In a local quantum field theory a dense set of Hilbert space
  vectors are constructed by applying polynomials of local field
  operators smeared with Schwartz test functions \cite{gelfand} in
  four space-time variables to the vacuum.  A pre-Hilbert space inner
  product of two of these states can be expressed as the vacuum
  expectation value of another polynomial in the smeared fields.  In
  general this space will have zero-norm vectors.  To pass to a
  Hilbert space representation it is necessary to replace vectors by
  equivalence classes of vectors that differ by zero norm vectors and
  add new vectors defined by Cauchy sequences of vectors to make the
  space complete.  There are standard methods to do this
  \cite{Wightman:1980}.  Any transformation that leaves these vacuum
  expectation values unchanged preserves this Hilbert space inner
  product and is by definition a unitary transformation.  The
  polynomials in the smeared fields are an algebra of operators; they
  are closed under addition, multiplication by complex numbers and
  operator products.  This algebra will be referred to as the local
  Heisenberg algebra.  The vacuum can be understood as a positive
  linear functional on this algebra.  Vacuum functionals that do not
  preserve this inner product lead to inequivalent Hilbert space
  representations of this algebra of operators.  Another algebra is
  the algebra of fields restricted to the light front, called the
  light-front Fock algebra.  It is defined by replacing fields smeared
  with 4-dimensional Schwartz test functions by test functions of the
  form $\delta (x^+) f(x^-,\mathbf{x}_{\perp})$ where
  $f(x^-,\mathbf{x}_{\perp})$ is a Schwartz test function in three
  variables with vanishing $x^-$ derivative \cite{Schlieder:1972qr}.
  The vacuum functionals for all free field theories are defined on
  this algebra and give the same vacuum expectation values.  For free
  fields there is a mapping from the local Heisenberg algebra to a
  {\it sub-algebra} of the light-front Fock algebra.  This maps
  polynomials of fields smeared with 4-dimensional Schwartz functions
  to polynomials of fields on the light front by enforcing the
  ``free-field mass shell'' condition to eliminate the $x^+$
  dependence.  This defines a sub-algebra of the light-front Fock
  algebra.  This will be referred to as the light-front mass $m$
  sub-algebra.  The test functions $f$ are mapped into functions,
  $\tilde{f}$, of light-front variables; these functions will be
  referred to as the light-front mass $m$ test functions.  If
  $\tilde{f}_i$ are the light-front mass $m$ test functions associated
  with the four dimensional Schwartz functions $f_i$ then \beq
  _{m_1}\langle 0 \vert \phi_{m_1}(f_1) \cdots \phi_{m_1} (f_n) \vert
  0 \rangle_{m_1} = _{m_1}\langle 0 \vert \phi_{m_1}(\tilde{f}_1)
  \cdots \phi_{m_1} (\tilde{f}_n) \vert 0 \rangle_{m_1} =
  _{m_2}\langle 0 \vert \phi_{m_2}(\tilde{f}_1) \cdots \phi_{m_2}
  (\tilde{f}_n) \vert 0 \rangle_{m_1}
\label{i:1}
\eeq
where the subscripts $m_1$ and $m_2$ refer to free fields and vacuum
functionals associated with masses $m_1$ and $m_2$.  This means that
free field theories with different masses are unitarily equivalent on
the light-front mass $m$ {sub-algebra}. Equation (\ref{i:1}) shows that
the Hilbert space inner product of the theory is equal to the vacuum
expectation value of elements of light-front mass $m$ sub-algebra
using free field theories of {\it any} mass.  In this case all of the
dynamical information is contained in the choice of light-front mass $m$
sub-algebra, rather than the vacuum.  The mapping from the Heisenberg
algebra to the light-front mass $m$ sub-algebra is not invertible
since it cannot distinguish four-dimensional Schwartz functions whose
Fourier transform agree on the mass shell.

Theories with different masses are associated with different sub-algebras.
Light-front vacuum expectation values for {\it different} sub-algebras
do not agree.  This resolves the problem of inequivalent representation
for free fields.

\item[] In the interacting case there is an irreducible algebra of
  asymptotic fields
  \cite{Haag:1958vt}\cite{Brenig:1959}\cite{Ruelle:1962}\cite{jost:1966}.
  These behave like free fields with physical masses
  \cite{strocchi}\cite{simon}, except the creation and annihilation
  operators transform $N$-particle scattering states to
  $N\pm 1$ particle scattering states.  Vacuum expectation
  values of polynomials of smeared asymptotic fields satisfy
  \beq
\langle 0 \vert \phi_{IN}(f_1) \cdots \phi_{IN} (f_n) \vert 0 \rangle
= \langle 0 \vert \phi_{OUT}({f}_1) \cdots \phi_{OUT}
({f}_n) \vert 0 \rangle = _{m_p}\langle 0 \vert
\phi_{m_p}({f}_1) \cdots \phi_{m_p} ({f}_n) \vert 0 \rangle_{m_P}
\label{i:2}
\eeq
where the fields in the third factor are free fields with {\it physical
masses} and the corresponding free-field vacuum.  The physics is in the
relation of the IN fields to the OUT fields or the IN or OUT fields
to the Heisenberg fields.  
It follows by combining (\ref{i:1}) and (\ref{i:2}) that the vacuum expectation values
of products of $IN$ fields are equal to light-front vacuum
expectation values of products of free fields smeared with
light-front mass $m_p$ test functions:
\beq
\langle 0 \vert \phi_{IN}({f}_1) \cdots \phi_{IN} ({f}_n) \vert 0 \rangle  =
_{m_p}\langle 0 \vert \phi_{m_p}(\tilde{f}_1) \cdots \phi_{m_p} (\tilde{f}_n) \vert 0 \rangle_{m_P} .
\label{i:3}
\eeq
Since the smeared IN fields are an irreducible set of Hilbert space
operators (assuming that the field theory is asymptotically complete),
smeared Heisenberg fields can be expanded in normal products of the IN
fields \cite{Haag:1955ev}\cite{Glaser:1957}\cite{Greenberg:1965}
\cite{nishijima}.
This means the vacuum expectation values of polynomials of smeared
interacting Heisenberg fields can be expressed as limits of vacuum
expectation values of polynomials in smeared IN fields which in turn
can be expressed as limits of vacuum expectation values of operators
in the light-front mass $m_P$ sub algebra.

The complication is that there are asymptotic fields associated
with each stable particle of the theory, which includes stable
composite systems, so the IN fields with different masses involve
mappings to different sub-algebras of the light-front Fock algebra.
The basic result is that the vacuum expectation
values of polynomials of smeared Heisenberg fields is equal to
light-front vacuum expectation values of elements of a sub-algebra of the
light-front Fock algebra.  While the interacting case is more
complicated, the result is the same as in the free-field case.  The
physics is in the sub-algebra of the light-front Fock algebra.  The
resulting light-front vacuum expectation value is independent of the
vacuum functional used to evaluate the vacuum expectation value.

For case of QCD the Heisenberg fields for quarks and gluons field
cannot be expanded in terms of the algebra asymptotic fields.  See
\cite{Greenberg:1978} for a possible way to introduce ``asymptotic
fields'' for confined particles.  An expansion in asymptotic fields is
not necessary for the vacuum expectation values involving smeared
quark and gluon Heisenberg fields to be expressible in terms of a
sub-algebra of quark and gluon fields in the light-front Fock algebra.
The unsolved problem is to find an appropriate sub algebra.

\item[2.] The physical vacuum expectation values of products of free
  fields can also be computed using products of suitably smeared
  canonical pairs of fields.  Formally there is a mapping from the
  Heisenberg field algebra to the algebra canonical fields
  restricted to a fixed time surface.  In this case the canonical
  field sub-algebras are different for different masses and,
  unlike the light-front case, the vacuum functionals on
  the sub-algebras are different.  They result in inequivalent
  representations of the canonical commutation relations.  In the
  light-front case the corresponding map to the light-front
  sub-algebra discussed in 1) carries all of the dynamical information
  of the theory and the inequivalence is due to the different
  sub-algebras rather that due to the different light-front vacuum
  functionals.
 
\item[3.] In the light-front case Noether's theorem on the light front
  results in expressions for all ten Poincar\'e generators expressed
  as elements of the irreducible light-front Fock algebra.  The
  mapping discussed in 1) maps Schwartz test functions in four
  variables to functions on the light front that vanish faster than
  any power of $p^+$ as $p^+ \to 0$.  If the Schwartz functions are
  restricted to the dense set of Schwartz functions having Fourier
  transforms with compact support, this dense set is mapped to a
  subspace of vectors on the light-front Fock space, and the power
  series expansion of $e^{-i P^-x^+/2}$ converges on this subspace of
  vectors. This means that the light front mass-m sub-algebra enforces
  boundary conditions that result in a well-defined initial valued
  problem.

  \item[4.] Formally the results of any dynamical calculation should
    be independent of the orientation of the light front. If the
    scattering operator is independent of the orientation of the light
    front then it is possible to use the scattering wave functions
    from both representations to construct a unitary representation of
    the rotation group.  It follows that rotational covariance is
    equivalent to independence on the orientation of the light front.
    This is not entirely trivial to achieve because it requires
    ensuring the rotational covariance of the asymptotic conditions,
    which is exactly the problem that appears when degenerate bound
    state solutions have the same magnetic quantum numbers.  Replacing
    the light-front Hamiltonian, $P^-$, by one of the transverse
    rotation generators has the property that all of the generators
    can be constructed using the commutation relations with the
    kinematic generators.  While this results in a consistent set of
    generators, it is still non-trivial because there are linear and
    non-linear constraints on the transverse rotation generator that
    come from Poincar\'e covariance.  What makes this problem
    difficult in Hamiltonian formulations of quantum field theory is
    that it is related to renormalization.  $p^+\to 0$ divergences
    with respect to one light front become $p^3\to \infty$ divergences
    using a different orientation of the light front.  Invariance
    under change of orientation of the light front implies that the
    renormalization of the $p^+\to 0$ and $\mathbf{p}\to \infty$
    divergences of the theory cannot be performed independently.  Note
    that restoring rotational covariance may not result in the same
    representation of the rotation group associated with the Noether
    charges, but the resulting theory will be unitarily equivalent to
    the one derived from Noether's theorem.  At the perturbative
    level this can be solved by appealing to the connection with
    covariant perturbation theory \cite{Mannheim:2020rod}, but how to
    achieve this at the non-perturbative level remains an open
    problem.

    \item[5.] The discussion of rotational invariance led to two
      important observations.  The first is that the full rotational
      covariance is not encoded in the dynamical $P^-$.  The second
      observation is the a proper renormalization of $P^-$ must be
      consistent with rotational covariance, which implies that the
      large $p$ and small $p^+$ divergences must be renormalized
      consistently.  Space reflection symmetry also relates these two
      kinds of divergences.  This means that the constraints on the
      renormalization of the $p^+\to 0$ and $p\to \infty$ divergences
      that are needed to define a theory consistent rotational
      covariance and space reflection symmetry do not come directly
      from the structure of $P^-$.  In perturbation theory the missing
      information can be determined by appealing to the covariant
      formulation of the theory.  In general the problem is difficult
      to separate from the problem of how to consistently renormalize all of the
      dynamical Poincar\'e generators outside of perturbation theory.

  \item[6.] The signal for spontaneous symmetry breaking is the
  presence of a 0 mass Goldstone boson.  Coleman \cite{coleman} gives the
  following condition for the presence of a 0 mass Goldstone boson
\beq
\lim_{R\to \infty}\int_{\vert \mathbf{x} \vert <R}d\mathbf{x}
\langle 0 \vert [j^0(x) ,\phi(y)] \vert 0 \rangle \not=0
\label{i:4}
\eeq
This is non-vanishing if the cutoff charge couples the vacuum to the
Goldstone boson.  This only assumes that the current is a local
operator valued distribution; this does not require the existence a
charge operator.  Central to the above argument is the locality
requirement that $[j^0(x) ,\phi(y)]=0$ for space-like separated $x$
and $y$ which cuts off the integral for sufficiently large $R$. This
argument fails on the light-front because separated points on the
light-front are not all space-like.  While this condition can in
principle be formulated on the light-front, in a dynamical theory the
structure of the light-front sub-algebra associated with an
interacting field is more complicated than the mass $m$ light-front
Fock algebra.  The symmetry breaking must involve the selection of a
particular sub-algebra of the light-front Fock algebra.

\end{itemize}  

These results are discussed in more detail in what follows.  The next
section reviews Poincar\'e invariance in quantum theories.  This
interpretation, due to Wigner \cite{Wigner:1939cj}, is relevant to
understanding Dirac's forms of dynamics which are relevant for
Hamiltonian formulations of quantum field theories and are reviewed in
section 3.  Inequivalent representations of the canonical commutation
relations are discussed in section 4. The irreducibility of covariant,
canonical and light-front Fock algebras is discussed in section 5.
This section also discusses the relation between the Heisenberg,
canonical, and light-front Fock algebras for free fields.  The
initial value problem is discussed in section 6.  The triviality of
the vacuum is discussed in section 7.  Rotational covariance is
discussed in section 8.  Comments on interacting theories are given in
section 9.  Spontaneous symmetry breaking is discussed in section
10. A summary and conclusion is given in section 11.  There are two
appendices.  The first summarizes the conventions used in this paper and
the second gives the form of the light-front Poincar\'e generators as
operators in the light-front Fock algebra.

\section{Poincar\'e Invariance and Hamiltonian formulations of relativistic  quantum theories}

The relativistic invariance of a quantum theory requires a unitary
(ray) representation, $U(\Lambda,a)$, of the component of the
Poincar\'e group connected to the identity \cite{Wigner:1939cj}.  Here
$(\Lambda,a)$ represents the semi-direct product of a Lorentz
transformation, $\Lambda$, followed by a spacetime translation by
a fixed four vector $a$.
This ensures that quantum probabilities, expectation values and
ensemble averages are independent of the inertial coordinate system.  The
Poincar\'e group is a 10 parameter group.  It is generated by four
one-parameter subgroups of space-time translations and six
one-parameter subgroups of Lorentz transformations.  The infinitesimal
generators of these unitary one-parameter subgroups are self-adjoint
operators \cite{Simon_I}.  They include the Hamiltonian, $P^0=H$, the
linear momentum, $\mathbf{P}$, the angular momentum, $\mathbf{J}$, and
the rotationless boost generators, $\mathbf{K}$.  The group
representation property
\beq U(\Lambda_2,a_2)U(\Lambda_1,a_1) =
U(\Lambda_2\Lambda_1,\Lambda_2 a_1 +a_2)
\label{p1}
\eeq
implies that the infinitesimal generators satisfy the
commutation relations:
\[
[P^{\mu},P^{\nu}]=0, \qquad [J^i, P^j]= i \epsilon^{ijk}P^k,
\qquad [J^i, J^j]= i \epsilon^{ijk}J^k,
\]
\[
[J^i, K^j]= i \epsilon^{ijk}K^k, \qquad
[K^i, K^j]= -i \epsilon^{ijk}J^k
\]
\beq
[K^i,P^i] = i\delta^{ij} H
\qquad
[K^i,H] = i P^i.
\label{p2}
\eeq
Light-front generators are linear combinations of these operators.
The relativistic analog of diagonalizing the Hamiltonian is to
decompose $U(\Lambda,a)$ into a direct integral of irreducible
representations.  This is equivalent to simultaneously diagonalizing the
mass and spin Casimir operators of the Lie algebra
\beq
M^2= (P^0)^2 - \mathbf{P}^2 \qquad \mbox{and} \qquad \mathbf{S}^2 = W^2 /M^2 
\eeq
where $W^{\mu}$ is the Pauli-Lubanski vector
\beq
W^{\mu}=(\mathbf{P}\cdot \mathbf{J}, H\mathbf{J}+ \mathbf{P}\times \mathbf{K}) .
\eeq
The transformation properties of states in each irreducible subspace is fixed by group theoretical considerations.

\section{Dirac's forms of Dynamics}

This section discusses Dirac's three forms of Hamiltonian dynamics,
and in particular the front-form of the dynamics.  The starting point
in discussing Dirac's forms of dynamics \cite{Dirac:1949cp} is the
assumption that the non-interacting and interacting theories are
formulated on the same representation of the Hilbert space as the
interacting theory.  While this is always true for quantum theories of
a finite number of particles, it is not generally true in for theories
with an infinite number of degrees of freedom \cite{polyzou:2021}.  It
is not even true for free field theories.  Dirac's forms of dynamics
are used in perturbative quantum field theory where the dynamics is
introduced as a perturbation of the non-interacting theory.  As long
as there are cutoffs both theories can be formulated on the same
Hilbert space representation.  As the cutoffs are removed the interacting theory passes to a different Hilbert space representation.  In this work
the existence of both an interacting and non-interacting unitary
representation of the Poincar\'e group that act on the same
representation of the Hilbert space will be assumed.

Given the assumption that the free and interacting unitary representations of
the Poincar\'e group act on the same representation of the Hilbert space,
the non-interacting representation of the Poincar\'e group 
also has a set of infinitesimal generators that are self-adjoint
operators on the same representation of the Hilbert space.  The
non-interacting generators are denoted with a ``0'' subscript:
\beq
\{ H_0, \mathbf{P}_0, \mathbf{J}_0,\mathbf{K}_0 \}.
\label{d1}
\eeq
For the linear momentum, $\mathbf{P}_0$, angular momentum
$\mathbf{J}_0,$ and rotationless boost generators, $\mathbf{K}_0$ the
spectrum of the operators is fixed by group representation properties
and is identical in the non-interacting and interacting theories.
These nine operators cannot be related by a single unitary
transformation because the commutator
\beq
[K^i,P^j] = i\delta^{ij} H 
\label{d2}
\eeq
would then require that $H$ and $H_0$ would be unitarily equivalent.
This is certainly not true if
$H$ has bound states.

Dirac pointed out that for closed sub-algebras that do not involve
the Hamiltonian it is possible to find a representation where all of
the operators in the sub-algebra are identical in the free and
interacting representations of the Poincar\'e group.  It follows from
the commutation relations (\ref{p2}) that two sub-algebras with this
property are the three-dimensional Euclidean algebra (instant
representation)
\beq
[P^{i},P^{j}]=0, \qquad [J^i, P^j]= i \epsilon^{ijk}P^k,
\qquad [J^i, J^j]= i \epsilon^{ijk}J^k,
\label{d3}
\eeq
and the Lorentz algebra (point representation)
\beq
[J^i, K^j]= i \epsilon^{ijk}K^k, \qquad
[K^i, K^j]= -i \epsilon^{ijk}J^k, \qquad
[J^i, J^j]= i \epsilon^{ijk}J^k .
\label{d4}
\eeq
Each of these algebras has 6 generators.  In the instant representation
$H\not=H_0$, and $\mathbf{K}\not=\mathbf{K}_0$.  In the point representation
$H\not=H_0$, and $\mathbf{P}\not=\mathbf{P}_0$.  Dirac identified
a third representation where the sub-algebra that generates the seven-parameter
subgroup of the Poincar\'e group that maps the light-front
hyperplane, $x^+=x^0+x^3=0$, to itself is the same in the non-interacting and
interacting theory.  In this case the non-interacting generators are the following linear combinations of the generators (\ref{p2}): 
\beq
P^1,P^2,P^+=P^0+P^3,J^3,K^3,\mathbf{E}_{\perp}:= \mathbf{K}_{\perp}- \hat{\mathbf{z}}
\times \mathbf{J}
\label{d5}
\eeq
while the generators
\beq
P^- = P^0-P^3 \not=P^-_0;
\qquad \mbox{and} \qquad \mathbf{F}_{\perp}:= \mathbf{K}_{\perp}+ \hat{\mathbf{z}}
\times \mathbf{J} \not= \mathbf{F}_{\perp 0} \qquad
\mbox{or} \qquad \mathbf{J}_{\perp} = 
\hat{\mathbf{z}} 
\times \mathbf{J} \not=\hat{\mathbf{z}} \times \mathbf{J}_{0\perp}  
\label{d6}
\eeq
involve interactions.  The dynamical generators can be taken as $P^-,F^1,F^2$ or
equivalently $P^-,J^1,J^2$.  The operators $K^3$ and $\mathbf{E}_{\perp}$, which generate
light-front preserving boosts, form a closed sub-algebra.

In all three of Dirac's forms of dynamics the
dynamical problem is to simultaneously diagonalize the mass $M$ and
spin $\mathbf{S}^2$ Casimr operators of the Poincare Lie algebra.
Sokolov and Shatnyi \cite{Sokolov:1977im}\cite{Keister:1996bd} established the equivalence
of these three representations of the Poincar\'e group when the
kinematic and dynamical representations of the Poincar\'e group
act on the same representation of the Hilbert space.

While Dirac's forms of dynamics are normally formulated in terms of
initial value surfaces, if the light front hypersurface,
$x^+=x^-+\hat{\mathbf{z}}\cdot \mathbf{x}=0$, is replaced by
$x^+=x^-+\hat{\mathbf{z}}\cdot \mathbf{x}=c\not=0$, the subgroup that
leaves the $x^+$-evolved surface invariant is only a six parameter
group.  However generators that are defined on the $x^+=0$ are all
that is needed.  This motivates the preference for discussing forms of
dynamics in terms of kinematic subgroups rather than initial value
surfaces.

\section{Inequivalent representations of the canonical commutation relations}

For a quantum theory of 1 degree of freedom any operator can be formally
expressed in the form
\beq
O = \int da db f(a,b)e^{iqa}e^{ipb} \qquad e^{iqa}e^{ipb} =e^{ipb}e^{iqa}e^{-iab}
\label{i1}
\eeq
where $p$ and $q$ are canonically conjugate momentum and coordinate operators
and $f(a,b)$ is a function of 2 real variables.
The operators $e^{iqa}$ and $e^{ipb}$ are an irreducible set of operator on
the Hilbert space of square integrable functions in $q$ (or $p$).  
Any operator that commutes with both of them is a constant multiple of
the identity. 

A key difference with canonical and light-front quantum theories is that while
both theories have irreducible sets of field operators, in canonical
field theories with different masses the vacuum functionals generate
inequivalent Hilbert space representations of the canonical
commutation relations, while in the light-front case there is a common
representation of the Weyl algebra (exponential form of the canonical
commutation relations (\ref{i1})) and both vacuum functionals agree on this
algebra.  General considerations are discussed in this section.  The
resolution is discussed in the following section.

The Stone Von Neumann's theorem
\cite{MHStone1}\cite{JvNeumann2}\cite{JvNeumann3} that demonstrated
the equivalence of the Schr\"odinger and Heisenberg pictures of
quantum mechanics breaks down for theories of an infinite number of
degree of freedom.  Haag \cite{Haag:1955ev} provided a simple example
that illustrates this breakdown for free fields.

For two quantum mechanical harmonic oscillators with different
frequencies, $\omega_i$,  the creation and annihilation operators are related to the
canonical coordinates and momenta by
\beq
q = {1 \over \sqrt{2\omega_i }}
\left ( a_i + a_i^{\dagger}  \right ) 
\qquad
p = -i \sqrt{{\omega_i \over 2}}
\left  (a_i - a_i^{\dagger}\right  ) 
\label{i2}
\eeq
where $\omega_i$ is the angular frequency of the $i^{th}$ oscillator.
Solving for one set of creation and annihilation operators in terms of the
other gives the canonical transformation
\beq
a_2 = \cosh(\eta) a_1  + \sinh (\eta)  a_1^{\dagger}
\label{i3}
\eeq
where $\eta$ is defined by 
\beq
\cosh (\eta) := {1 \over 2} \left (\sqrt{{\omega_{2} \over
\omega_{1}}} +
\sqrt{{\omega_{1}\over \omega_{2} }} \right )
\qquad
\sinh (\eta) := {1 \over 2} \left ( \sqrt{{\omega_{2} \over \omega_{1}}} -
\sqrt{{\omega_{1}\over \omega_{2} }} \right ) .
\label{i5}
\eeq
This canonical transformation can be realized
by a unitary transformation $e^{iG}$
with infinitesimal generator 
\beq
G = - {i\over 2} \eta (a_1  a_1
 - a_1^{\dagger} a_1^{\dagger}  ) .
\label{i7}
\eeq
The ground state vectors of the two oscillators are related by this unitary transformation
\beq
\vert 0 \rangle_2 = e^{iG}\vert 0 \rangle_1 .
\label{i6}
\eeq

For free scalar fields with different masses,
the canonical coordinates and momenta are replaced by the fields $\phi(0,\mathbf{x})$ 
and
$\pi(0,\mathbf{x})$.  They can be expressed in terms of
creation and annihilation operators by
\beq
\phi (x) = {1 \over (2 \pi)^{3/2}}
\int {d\mathbf{p} \over \sqrt{2\omega_{m_i}(\mathbf{p})}}
\left  (e^{i p\cdot x} a_i(\mathbf{p}) +
e^{-i p\cdot x} a_i^{\dagger}(\mathbf{p}) \right) 
\label{i8}
\eeq
\beq
\pi (x) = -{i \over (2 \pi)^{3/2}}
\int d\mathbf{p} \sqrt{{\omega_{m_i}(\mathbf{p}) \over 2}}
\left  (e^{i p\cdot x} a_i(\mathbf{p}) -
e^{-i p\cdot x} a_i^{\dagger}(\mathbf{p}) \right) 
\label{i9}
\eeq
where $\omega_m(\mathbf{p}):= \sqrt{m^2+\mathbf{p}^2}$ is the energy of a particle of mass $m$.

Following the construction in the one degree of freedom case,
the creation and annihilation operators in fields $\phi_m(0,\mathbf{x})$
with different masses are
related by a canonical transformation of the form
\beq
a_2(\mathbf{p}) = \cosh(\eta (\mathbf{p})) a_1(\mathbf{p})  + \sinh (\eta (\mathbf{p}) ) a_1^{\dagger}(\mathbf{p})
\label{i10}
\eeq
where
\beq
\cosh (\eta(\mathbf{p})) := {1 \over 2} \left (\sqrt{{\omega_{m_2}(\mathbf{p}) \over
\omega_{m_1}(\mathbf{p})}} +
\sqrt{{\omega_{m_1}(\mathbf{p})\over \omega_{m_2}(\mathbf{p} }} \right )
\label{i11}
\eeq
\beq
\sinh (\eta(\mathbf{p})) := {1 \over 2} \left ( \sqrt{{\omega_{m_2}(\mathbf{p}) \over \omega_{m_1}(\mathbf{p})}} -
\sqrt{{\omega_{m_1}(\mathbf{p})\over \omega_{m_2}(\mathbf{p} }} \right ) .
\label{i12}
\eeq

If this canonical transformation was implemented by a unitary
transformation the generator, $G$, of the unitary operator, $U=e^{iG}$,
would be
\beq
G = - {i\over 2} \int \eta (\mathbf{p}) (a_1 (\mathbf{p}) a_1
(\mathbf{p}) - a_1^{\dagger}(\mathbf{p}) a_1^{\dagger}(\mathbf{p}) )d
\mathbf{p} 
\label{i13}
\eeq
and the vacuum vectors in the two theories would be related by
\beq \vert 0 \rangle_2 = e^{iG}\vert 0 \rangle_1,
\label{i14}
\eeq
however a straightforward calculation gives
\beq
\Vert G \vert 0 \rangle_1 \Vert^2 = {1 \over 4} \int \eta(\mathbf{p})^2
d\mathbf{p} \delta (0)= \infty .
\label{i15}
\eeq
Since the norm of any vector is a linear combination of products of
contractions multiplied by $_2\langle 0 \vert 0 \rangle_2$ the
generator $G$ has an empty domain on the Hilbert space generated by
the vacuum $a_1(\mathbf{p})\vert 0 \rangle_1=0$ of $\phi_1(,\mathbf{x}),
\pi_1(0,\mathbf{x})$.

The conclusion is that even for free canonical field theories the
vacuum vectors for theories with different masses live in inequivalent
representations of the Hilbert space.

In the canonical case, assuming that the Hamiltonian is quadratic in
the generalized momentum, the vacuum actually determines the
Hamiltonian \cite{Araki:1964}\cite{PhysRev.117.1137}.  This also
follows because while the fields for any mass satisfy the same
canonical commutation relations, the choice $a_i(\mathbf{p}) \vert 0
\rangle_i=0$ of vacuum determines the mass.  This is in contrast to
the light-front case where vacuum expectation values of fields
on the light front have no dynamical information.

Another way to understand this inequivalence is to recall that 
in quantum field theory the vacuum is a linear functional on an algebra
of field operators; where this algebra is generated by fields smeared
with Schwartz \cite{gelfand} test functions in all four spacetime variables.
For free fields the inner
product of two one-particle states has the form
\beq
\langle f \vert g \rangle = \sum_{ij}
\int f_i^*(x) \langle 0 \vert \phi_i^{\dagger} (x) \phi_j(y) \vert 0 \rangle
g_j(y) d^4x d^4y.
\label{i16}
\eeq
The Fourier transform of $\langle 0 \vert \phi_i^{\dagger} (x) \phi_j(y) \vert 0 \rangle$ is the product of a
spinor matrix \cite{Wightman:1980}, a mass-shell delta function and a Heaviside
function that selects the positive-energy branch of the mass shell.
This means that two test functions with Fourier transforms that agree
on the mass shell represent the same vector, so Hilbert space vectors
are represented by equivalence classes of test functions.  If the mass
is changed, two functions in the same equivalence class with respect to the
first mass do not necessarily have Fourier
transforms in the same equivalence class with respect to
the second mass.  This means that there is
no correspondence of equivalence classes associated with one mass with
equivalence classes associated with a second mass.
For free fields all Wightman functions are products of two point functions.
Given the same field algebra, different two point functions are determined by the vacuum
functional that gives the Wightman functions.

The resolution of the vacuum problem is alluded to in
\cite{Schlieder:1972qr}.  In general a given theory has one vacuum,
which is a linear functional on an algebra of local Heisenberg fields.
The algebra can be generated polynomials of 
fields
smeared with Schwartz functions in 3+1 variables or
linear combinations of bounded functions of fields
smeared with Schwartz functions in 3+1 variables:
\[
e^{i \phi (f_i)}.
\]
It is also possible that a given vacuum functional may still be
defined by taking limits that put the support of the test functions
on different hypersurfaces of Minkowski space.  
What is pointed out in \cite{Schlieder:1972qr} is that when the
algebra of free field theories with different masses is restricted to
fields smeared with test functions that give the same value to two
point Wightman functions, then the theories defined on this
sub-algebra become unitarily equivalent.  In this case the different
vacuum functionals agree on this sub-algebra, but they do not agree
when the fields are smeared with arbitrary Schwartz functions.  This
limited algebra does not have enough operators to distinguish the two
theories.

This suggests that while different field theories have different
vacuum functionals, the vacuum functionals may agree when the space of
test functions is limited to functions supported on the light front,
although in the light-front case the different vacuum functionals are
only defined for fields smeared with test functions supported on the
light front with Fourier transforms that vanish at $p^+=0$, which
corresponds to infinite momentum \cite{Schlieder:1972qr}.  These will
be referred to as Schlieder-Seiler test functions.

In the light-front case, for free fields, the condition $\int
a(\tilde{\mathbf{p}})f(\tilde{\mathbf{p}})d\tilde{\mathbf{p}} \vert 0
\rangle_{LF} =0$, where $\tilde{\mathbf{p}}:= (p^+,p^1,p^2)$ and
$f(\tilde{\mathbf{p}})$ is a Schlieder-Seiler test function, does not
have enough test functions to 
distinguish vacuum functionals for different free field theories.  The
spaces generated by applying elements of this algebra to different
vacua are unitarily equivalent \cite{Schlieder:1972qr}.

The conclusion is that while each field theory will have a different
vacuum vector, defined as a linear functional on the Heisenberg field
algebra, different vacuum functionals can agree on a sub-algebra.  For
the case of free fields the test functions for the light front and
canonical field algebras are not sub-algebras of the Heisenberg
algebra because the test functions are not four dimensional Schwartz
functions, but the vacuum functionals are defined on
three-dimensional hypersurfaces
because the vacuum functionals are only sensitive to the values of the
Fourier transforms of the test functions on the mass shells, so the
different vacuum functionals can still make sense on these algebras.
This will be shown explicitly in the next section.

\section{Irreducibility} 

In this section it is shown that the Hilbert space generated by the
vacuum and the local Heisenberg algebra of free field operators of a
given mass can alternatively represented by the vacuum and an algebra
of canonical fields at fixed time or the vacuum and an algebra of
fields on the light front.  It follows that vacuum expectation values
of elements of a sub-algebra of fields on the light front can be used
to construct the Wightman functions of the field theory.  These
Wightman functions can also be constructed using the algebra of
canonical fields restricted to a fixed time surface.  For free fields
this demonstrates the equivalence of all three representations.

To show this the irreducibility of the canonical or
light-front fields is used to express creation and annihilation operators in terms of
fields on the light front or fixed time surface.  In this way the
Heisenberg field can be expressed in terms of fields restricted to the
light front or a fixed time surface.  With this connection the
covariant Poincar\'e generators get mapped to light-front
generators that act on a Hilbert space generated by applying a
(sub)algebra of fields restricted to light-front
hyperplanes to the vacuum.

The canonical algebra is generated by polynomials in
\beq
\phi(f) = \int d\mathbf{x}\phi (t=0,\mathbf{x}) f(\mathbf{x})
\qquad
\pi(f) = \int d\mathbf{x}\pi (t=0,\mathbf{x}) f(\mathbf{x})
\qquad
f(\mathbf{x}) \in S(\mathbb{R}^3)
\label{ir00}
\eeq
where the $f(\mathbf{x})$'s are Schwartz functions.

The light-front Fock algebra is generated by polynomials in
\beq
\phi(f) = \int {dx^-d^2\mathbf{x}_{\perp} \over 2} \phi (x^+=0,x^- ,
\mathbf{x}_{\perp})f (x^-,
\mathbf{x}_{\perp})
\qquad
f((x^-,
\mathbf{x}_{\perp}) \in S(\mathbb{R}^3) \qquad {\partial f \over \partial x^-}=0
.
\label{ir01}
\eeq
The test functions in (\ref{ir01}) are
Schlieder-Seiler functions.

Normally irreducibility is discussed in terms of the exponential form (\ref{i1}) 
of the canonical commutation relations which avoids the use of unbounded
operators, but for practical purposes what is needed is to be able to
separately extract the creation and annihilation operators from the
field operators, since any operator can be expressed in terms of creation and annihilation operators.

The creation and annihilation operators for free fields restricted to a fixed-time surface can be extracted from the fields (\ref{i8}) and the generalized momenta
(\ref{i9}) by taking Fourier transforms:
\beq
a(\mathbf{p})= {1 \over \sqrt{2\omega_{m_i}(\mathbf{p})}}
(\omega_{m_i}(\mathbf{p}) \hat{\phi}(\mathbf{p})_{x^0=0}  +i \hat{\pi}(-\mathbf{p})_{x^0=0}),
\label{ir2}
\eeq
\beq
a^{\dagger} (\mathbf{p})= {1 \over \sqrt{2\omega_{m_i}(\mathbf{p})}}
(\omega_{m_i}(\mathbf{p}) \hat{\phi}(\mathbf{p})_{x^0=0}  -i \hat{\pi}(-\mathbf{p})_{x^0=0} ),
\label{ir2a}
\eeq
where
\beq
\hat{\phi}(\mathbf{p})_{x^0=0} :=
\int  {d\mathbf{x} \over (2 \pi)^{3/2}}e^{-i \mathbf{p}\cdot \mathbf{x}}
\phi (x^0=0,\mathbf{x})
\qquad
\hat{\pi}(\mathbf{p})_{x^0=0}  :=
\int  {d\mathbf{x} \over (2 \pi)^{3/2}}e^{-i \mathbf{p}\cdot \mathbf{x}}
\pi(x^0=0,\mathbf{x})
\label{ir3}
\eeq
are Fourier transforms of the fields at time $t=0$ 
and $\omega_m(\mathbf{p})= \sqrt{m^2 + \mathbf{p}^2}$ is the energy.
The operators $a (\mathbf{p})$ and $a^{\dagger} (\mathbf{p})$ are
operator valued distributions.  In this case the linear combination of the field operators that give the creation and annihilation operators depend on the mass.
 
The free Heisenberg field (\ref{i8}) can also be expressed in light-front variables
by replacing integrals over the light front components of the momentum, 
$\mathbf{p}$ in (\ref{i8}) by integrals over $\tilde{\mathbf{p}}:=(p^+, \mathbf{p}_{\perp})$.  Using
\beq
\vert { \partial (\tilde{\mathbf{p}})\over \partial (\mathbf{p})}\vert
 = {p^+ \over \omega_m (\mathbf{p})}
\qquad
a(\tilde{\mathbf{p}}) :=  a(\mathbf{p})
\sqrt{{\omega_m (\mathbf{p})\over p^+}}
\label{ir4}
\eeq
and
\beq
[a (\mathbf{p}),a^{\dagger} (\mathbf{p}']= \delta (\mathbf{p}-\mathbf{p}')
\eeq
gives
\beq
[a (\tilde{\mathbf{p}}),a^{\dagger} (\tilde{\mathbf{p}}']=
\delta (\tilde{\mathbf{p}}-\tilde{\mathbf{p}}')=
\delta (\mathbf{p}_{\perp}-\mathbf{p}_{\perp}') \delta (p^+-p^{+\prime})
\label{ir5}
\eeq
and the following expression for the field in terms of light-front
variables
\beq
\phi (x) = {1 \over (2 \pi)^{3/2}}
\int {d\mathbf{p}_{\perp} dp^+ \theta (p^+) \over \sqrt{2p^+}}
\left  (e^{i p\cdot x} a(\tilde{\mathbf{p}}) +
e^{-i p\cdot x} a^{\dagger}(\tilde{\mathbf{p}}) \right)
\label{ir6}
\eeq
where
\beq
p^- = {m^2 +\mathbf{p}_{\perp}^2 \over p^+}
\qquad
p\cdot x = -{1 \over 2}(p^+x^-+p^-x^+) + \mathbf{p}_{\perp}\cdot
\mathbf{x}_{\perp}.
\label{ir7}
\eeq
Both (\ref{ir6}) and (\ref{i8}) are different ways of writing the {\it same} field operator.
The Fourier transform of the field (\ref{ir6}) restricted to the light front is
\beq
\hat{\phi} (\tilde{\mathbf{p}})_{x^+=0} :=
{1 \over (2 \pi)^{3/2}} \int {d\mathbf{x}_{\perp}dx^- \over 2}
\phi (\mathbf{x}_{\perp},x^-,x^+=0)e^{i ( {p^+ x^- \over 2}- \mathbf{p}_{\perp}\cdot \mathbf{x}_{\perp}) } =
\label{ir8}
\eeq
\beq
{1 \over \sqrt{2p^+}}\theta (p^+) a(\tilde{\mathbf{p}})+
{1 \over \sqrt{-2p^+}}\theta (-p^+) a^{\dagger}(-\tilde{\mathbf{p}}) 
\label{ir9}
\eeq
which express both the creation and annihilation operators 
\beq
a(\tilde{\mathbf{p}}) = \sqrt{2p^+}\theta (p^+) \hat{\phi} (\tilde{\mathbf{p}})_{x^+=0}
\qquad
a^{\dagger}(\tilde{\mathbf{p}}) = \sqrt{2p^+}\theta (p^+) \hat{\phi} (-\tilde{\mathbf{p}})_{x^+=0}
\label{ir10}
\eeq
in terms of the field restricted to the light front, without using any
information normal to the light front.  This is in contrast to
the canonical case where $\pi(x)= \dot{\phi}(x)$ involves a derivative
normal to the $t=constant$ surface. 

Since the free local Heisenberg field can be expressed in terms
of creation annihilation operators and the creation and annihilation
operators can be expressed in terms of either
$\phi (t=0,\mathbf{x})$ and  $\pi (t=0,\mathbf{x})$
or $\phi (x^+=0,x^-, \mathbf{x}_{\perp})$
it follows that the Heisenberg field $\phi(x)$ can be expressed 
in terms of $\phi (t=0,\mathbf{x})$ and  $\pi (t=0,\mathbf{x})$
or $\phi (x^+=0,x^- ,\mathbf{x}_{\perp})$.

A dense set of vectors in the physical Hilbert space of a quantum field
theory is constructed by applying polynomials, $A,B,\cdots $ of smeared Heisenberg field
operators of the form
\beq
\phi (f) := \int d^4 x\sum_i \phi_i (x) f_i (x),
\label{ir11}
\eeq
to the vacuum of the theory, 
where $f_i (x)$ are 4-variable Schwartz functions.  The physical vacuum is a
positive linear functional $L$ on this algebra of polynomials.
Vectors are be represented by elements
$A,B$ of this algebra with inner product
\beq
\langle A \vert B \rangle =  L(A^{\dagger}B) =
\langle 0 \vert A^{\dagger}B \vert 0 \rangle
\label{ir12}
\eeq
(after eliminating zero norm vectors and including Cauchy sequences).

In both the canonical and light-front cases there are also algebras
generated by operators of the form
\beq
\phi_C({f}) := \int d^3 x\sum_i \phi_i (t=0,\mathbf{x}) f_i (\mathbf{x})
\qquad 
\pi_C(f) := \int d^3 x\sum_i \pi_i (t=0,\mathbf{x}) f_i (\mathbf{x})
\label{ir13}
\eeq
and
\beq
\phi_{LF}({f}) := \int {d^2\mathbf{x}_{\perp}dx^- \over 2} \sum_i \phi_i (x^+=0,x^-,\mathbf{x}_{\perp})
f_i (x^-, \mathbf{x}_\perp).
\label{ir14}
\eeq
where in the canonical case $f_i$ are 3-variable Schwartz test functions
and in the light-front case the test functions are 3-variable Schwartz functions
with vanishing $x^-$ derivative \cite{Schlieder:1972qr} .

While the algebras generated by polynomials in the smeared fields in
(\ref{ir13}) and (\ref{ir14}) are distinct and different from the
local Heisenberg field algebra generated by the smeared fields in
(\ref{ir11}), for free fields any $\phi(f)$ in (\ref{ir11}) can be
expressed as a linear combination of the fields $\phi_C({g})$ and
$\pi_C({g})$ or $\phi_{LF}(g)$ for suitable test functions $g$ related to $f$.  In
addition the vacuum expectation values (\ref{ir12}) of elements in the
Heisenberg field algebra can also be expressed as vacuum expectation values
of operators in the canonical or light-front field algebras.

While (for the case of free fields) vacuum functionals for different
masses are different, they become unitarily equivalent
\cite{Schlieder:1972qr} when they
are applied to elements of the light-front Fock algebra.
The equivalence follows from
\[
_1\langle 0 \vert \phi_1 (f_1,x^+=0) \cdots \phi_1 (f_n,x^+=0) \vert 0 \rangle_1=
_2\langle 0 \vert \phi_2 (f_1,x^+=0) \cdots \phi_2 (f_n,x^+=0) \vert 0 \rangle_2
\]
for all $f_n$ test functions on the light front with vanishing $x^-$ derivative.

Next it is shown that the algebra of free Heisenberg fields can be
mapped to the canonical or light front field algebras.  In this way
vacuum expectation values of operators in the Heisenberg
field algebra can be expressed as vacuum expectation values of
operators in either the canonical or light front field algebras.

To show this use (\ref{ir2}) and (\ref{ir2a}) in
(\ref{i8}):
\[
\phi(f) = \int f (x) \phi(x) d^4 x =
\]
\[
\int d^4 x f(x) \int d\mathbf{y} d\mathbf{p} 
{1 \over  (2\pi)^3 } \left (
\cos (\omega_m (\mathbf{p})x^0)e^{i \mathbf{p}\cdot (\mathbf{x}-\mathbf{y})}\phi(y^0=0,\mathbf{y}) + 
{\sin (\omega_m (\mathbf{p})x^0) \over \omega_m (\mathbf{p})}
e^{i \mathbf{p}\cdot (\mathbf{x}-\mathbf{y})}\pi(y^0=0,\mathbf{y})
\right ) =
\]

\beq
\int d^4x f (x) \int  d\mathbf{y} [ {\cal K}_{\phi m} (x,\mathbf{y}) \phi(y^0=0,\mathbf{y}) +
{\cal K}_{\pi m} (x,\mathbf{y})\pi(y^0=0,\mathbf{y}) ]
\label{ir15}
\eeq
which defines the kernels ${\cal K}_{\phi m}$ and ${\cal K}_{\pi m}$.
This can be expressed as
\beq
\phi(f) = \int d\mathbf{y}( f_{\phi} (\mathbf{y}) \phi_C(\mathbf{y})
+ f_{\pi} (\mathbf{y}) \pi_C(\mathbf{y}))
\label{ir16}
\eeq
where
\beq
f_{\phi} (\mathbf{y})= \int d^4 x f(x) {\cal K}_{\phi m} (x,\mathbf{y})
\qquad
f_{\pi} (\mathbf{y})= \int d^4 x f(x) {\cal K}_{\pi m} (x,\mathbf{y})
\label{ir17}
\eeq
with similar relations for $\pi(f)$.  This means that the algebra of
operators generated by smearing the free Heisenberg field with
Schwartz functions in four spacetime variables can be expressed in
terms of fixed time canonical pairs of fields smeared with 
Schwartz functions in three variables. This is because the mass-shell
condition eliminates one of the variables.

This exercise can be repeated in the light-front case. 
In the light-front case the vacuum is not sensitive to the
dynamics when applied to operators in the light-front sub-algebra. 

The key observation in the light-front case is that the Heisenberg
algebra is mapped to a {\it proper sub-algebra} of the light-front Fock algebra and
the dynamical information is contained in the sub-algebra.  A smeared Heisenberg field has
the form
\[
\phi(f) = \int f (x) \phi(x) d^4 x =
\]
\[
\int d^4x f(x) \int {d \tilde{\mathbf{p}} \theta (p^+) \over (2 \pi)^{3/2}\sqrt{2 p^+}}
\left [e^{ip\cdot x} a(\tilde{\mathbf{p}})+
e^{-ip\cdot x} a^{\dagger}(\tilde{\mathbf{p}}) \right ]. 
\]
Inserting the expressions (\ref{ir10}) for the light-front creation and annihilation operator in terms of fields restricted to the light front gives 
\[
=\int d^4 x f(x) { d \tilde{\mathbf{p}} \theta (p^+) \over (2 \pi)^{3/2}\sqrt{2 p^+}}
\left [e^{ip\cdot x} \sqrt{2p^+} \theta (p^+)
\hat{\phi}(\tilde{\mathbf{p}})_{\vert x^+=0}+
e^{-ip\cdot x}  \sqrt{2p^+} \theta (p^+)
\hat{\phi}(-\tilde{\mathbf{p}})_{\vert x^+=0}
\right ]= 
\]
\[
\int d^4x {f(x) d \tilde{\mathbf{p}} \theta (p^+) \over (2 \pi)^{3}}
\left [e^{ip\cdot x} e^{i {p^+ y^-\over 2} - i\mathbf{p}_{\perp}\cdot
\mathbf{y}_{\perp}}  +
e^{-ip\cdot x}  e^{-i {p^+ y^-\over 2} + i\mathbf{p}_{\perp}\cdot
\mathbf{y}_{\perp}}
\right ] {dy^+ d\mathbf{y}_{\perp} \over 2} \phi(\tilde{\mathbf{y}})_{\vert x^+=0}
= 
\]
\beq
\int d^4x f(x) { d \tilde{\mathbf{p}} \theta (p^+) \over (2 \pi)^{3}}
\left [e^{-i x^+ {\mathbf{p}_{\perp}^2+m^2 \over 2p^+}- 
i {p^+\cdot (x^--y^-)\over 2} + i\mathbf{p}_{\perp}\cdot
(\mathbf{x}_{\perp}- \mathbf{y}_{\perp})}  +
e^{i x^+ {\mathbf{p}_{\perp}^2+m^2 \over 2p^+}+i {p^+\cdot (x^--y^-)\over 2} - i\mathbf{p}_{\perp}\cdot
(\mathbf{x}_{\perp} - \mathbf{y}_{\perp})}
\right ] {dy^+ d\mathbf{y}_{\perp} \over 2} \phi(\tilde{\mathbf{y}})_{\vert x^+=0}
. 
\label{ir18}
\eeq
This expression defines a kernel:
\beq
{\cal K}_m(x,\tilde{\mathbf{y}}):=
\int {d \tilde{\mathbf{p}} \theta (p^+) \over (2 \pi)^{3}}
\left [e^{-i x^+ {\mathbf{p}_{\perp}^2+m^2 \over 2p^+}- 
i {p^+\cdot (x^--y^-)\over 2} + i\mathbf{p}_{\perp}\cdot
(\mathbf{x}_{\perp}- \mathbf{y}_{\perp})}  +
e^{i x^+ {\mathbf{p}_{\perp}^2+m^2 \over 2p^+}+i {p^+\cdot (x^--y^-)\over 2} - i\mathbf{p}_{\perp}\cdot
(\mathbf{x}_{\perp} - \mathbf{y}_{\perp})}
\right ]
\label{ir19}
\eeq
that is used to define the mass $m$ light-front test function
$f_m(\tilde{\mathbf{y}})$
associated with a four-variable Schwartz function $f(x)$:
\beq
f_m (\tilde{\mathbf{y}}) = \int d^4 x f(x) {\cal K}_m (x ,\tilde{\mathbf{y}})=
{\cal K}_m (f ,\tilde{\mathbf{y}}).
\label{ir20}
\eeq
Changing the sign of $\tilde{\mathbf{p}}$ in the second term in
(\ref{ir19}) gives
the following expression for the kernel 
\beq
{\cal K}_m(x,\tilde{\mathbf{y}}):=
\int {d \tilde{\mathbf{p}} \over (2 \pi)^{3}}
e^{-i x^+ {\mathbf{p}_{\perp}^2+m^2 \over 2p^+}} 
 e^{-i {p^+\cdot (x^--y^-)\over 2} + i\mathbf{p}_{\perp}\cdot
(\mathbf{x}_{\perp}- \mathbf{y}_{\perp})}.
\label{ir21}
\eeq
In both (\ref{ir15}) and (\ref{ir21}) if the Klein Gordon operator, $(\square^2 -m^2)$, is applied to
a test function in $S(\mathbb{R}^4)$, integrating by parts and applying the Klein Gordon
operator to the kernels in  (\ref{ir15}) and (\ref{ir21}) gives $0$.
In these expressions the dynamics enters through these kernels, which satisfy the field equations.

These kernels pick out the part of $f(x)$ on the mass shell.
These mappings are not invertible since there are many test functions
$f(x)\in S(\mathbb{R}^4)$ whose Fourier transforms agree on the mass shell.

The expression for the smeared Heisenberg field becomes
\beq
\phi(f) = \int {dy^+ d\mathbf{y}_{\perp} \over 2}
f_m ( \tilde{\mathbf{y}}) \phi(\tilde{\mathbf{y}})_{\vert y^+=0}
:=
\phi_{LF}(f_m)
\label{ir22}
\eeq
which demonstrates that operators in the Heisenberg field algebra can be expressed
as operators in the light-front Fock algebra.  

Equation (\ref{ir20}-\ref{ir21}) can be used to find the Fourier transform
$\hat{f}_m( \tilde{\mathbf{p}})$
of
$f_m(\tilde{\mathbf{y}})$ assuming that $f(x)$ in the Heisenberg field
is a real Schwartz function and $\hat{f}(p)=\hat{f}(p^-,p^+,\mathbf{p}_{\perp})$ is its Fourier transform:
\beq
\hat{f}_m( \tilde{\mathbf{p}}) = 
\sqrt{2 \pi} \hat{f}({\mathbf{p}_{\perp}^2 + m^2 \over p^+},\tilde{\mathbf{p}}).
\label{ir23}
\eeq   
It
follows from (\ref{ir23}) for $m\not=0$ that
$\hat{f}_m(
\tilde{\mathbf{p}})$ vanishes faster than any power of $p^+$ as
$p^+\to 0$.  This is because $\hat{f}(p)$ is also a Schwartz function
which vanishes faster that any polynomial in $1/p^-$ for large $p^-$.

This means that the Fourier transforms of the light-front
mass $m$ test functions, (\ref{ir20}), vanish for light-like
light-front momenta. It also ensures that fields on the light front
smeared with the $f_m(\tilde{\mathbf{y}})$ generate a proper sub-algebra of the
light-front field algebra.

Another observation is
that the kernel ${\cal K}_m$ in the mixed
coordinate-momentum representation
\beq
{\cal K}_m (x ,\tilde{\mathbf{p}}):=
\int {\cal K}_m (x ,\tilde{\mathbf{y}})
     {dy^+ d\mathbf{y}_{\perp} \over 2} e^{i \tilde{\mathbf{y}} \cdot \tilde{\mathbf{p}}}
\label{ir23a}
\eeq
maps the covariant $x^+$ translation generator 
\beq
2 i {\partial \over \partial x^+} {\cal K}_m (x ,\tilde{\mathbf{p}})=  
{\cal K}_m (x ,\tilde{\mathbf{p}}) {\mathbf{p}^2_{\perp}+m^2 \over p^+}
\label{ir24}
\eeq
to the
light-front Hamiltonian in the light-front Fock algebra.
More
generally the kernel ${\cal K}_m$ maps the covariant (differential)
form of the Poincar\'e Lie algebra to the light-front representation
of the Poincar\'e Lie algebra.

In the light-front case the smeared Heisenberg fields (\ref{ir11}) can
be expressed in terms of fields on the light front smeared with
light-front mass $m$ test functions.  Polynomials in the smeared
Heisenberg fields (\ref{ir11}) applied to the  vacuum can be
expressed as polynomials in the corresponding light-front fields
smeared with  light-front mass $m$ test functions applied to 
vacuum.   Finally vacuum expectation values of
products of smeared Heisenberg fields can be expressed as
vacuum expectation values of operators in a sub-algebra 
of the light-front field algebra:
\beq
\langle 0 \vert \phi(f) \phi(g) \vert 0 \rangle =
\langle 0 \vert \phi_{LF}(f_m) \phi_{LF}(g_m) \vert 0 \rangle
\eeq
For free fields the Wightman functions are products of
two-point functions, so this immediately generalizes to all Wightman functions.

The image of the Heisenberg field algebra under
this mapping will be called the mass $m$ light-front Fock algebra.
It is a sub-algebra of the light-front Fock algebra.

Note that while the different free-field vacuum functionals are
unitarily equivalent when restricted to the light-front Fock algebra,
because the mappings from the Heisenberg algebras to the light front
sub-algebras are not invertible, this equivalence does not extend to
the canonical or Heisenberg algebras.  It does however indicate that
any one of the unitarily equivalent vacua on the light front can be
used with the mass m light front field algebra to construct the
free-field Wightman
functions.  This identification is in the following sense:
\[
_1\langle 0 \vert \phi_{1LF}(f_{1m})\cdots  \phi_{1LF}(f_{nm}) \vert 0 \rangle_1
=_2\langle 0 \vert \phi_{2LF}(f_{1m}) \cdots \phi_{2LF}(f_{nm}) \vert 0 \rangle_2
\]

The conclusion is that for free fields the Heisenberg field algebra
can be mapped to a sub-algebra of the canonical or light-front Fock
algebra.  While free field theories are related by unitary
transformations when the algebra is restricted to the light-front
Fock algebra, the sub-algebras associated with different theories are
different and the mappings ${\cal K}_m$ are not invertible, so it is
not possible to use this unitary equivalence to find unitary
transformations that relate free field theories with different masses.

It is useful to summarize these observations:
\begin{itemize}

\item [1.] For free fields the vacuum functional of each theory is 
  defined on the light-front Fock algebra.  The different vacuum
  functions are unitarily equivalent when restricted to that
  algebra.

\item [2.] The light-front Fock space is equivalent to a generic
  Hilbert space of square integrable functions of light-front
  variables that vanish at $p^+=0$.  Creation and annihilation
  operators can be defined that create and annihilate ``partons'' that
  are not associated with any mass.  The identification with particles
  comes from restricting the light-front Fock space to a subspace
  generated by a sub-algebra of the light-front Fock algebra applied
  to a generic light-front vacuum.
  
\item [3.] The dynamics enters through a mapping from the Heisenberg
  field algebra to a sub-algebra of the light-front Fock algebra.  The
  Wightman functions of the field theory can be constructed evaluating
  products of the light-front fields smeared with light-front mass $m$
  test functions using any free-field vacuum functional.
  The kernel ${\cal K}_m$, which has the dynamical content of the theory,
  is a solution the Klein -Gordon equation with mass $m$.
  
\item [4.] Because the mappings ${\cal K}_m$ are not invertible, this
  equivalence does not extend to the Heisenberg algebra.  Specifically
  if A denotes the Heisenberg algebra, $K_mA$ denotes the mass $m$
  light-front algebra and $L_m$ denotes the mass $m$ vacuum functional then
\[  
L_{m_1}[A]= L_{m_1}[K_{m_1}A]=L_{m_2}[K_{m_1}A]  
\]
\[
L_{m_1}[A]\not= L_{m_2}[A] \qquad L_{m_1}[K_{m_1}A]\not=L_{m_1}[K_{m_2}A]
\]
\[
L_{m_2}[A]= L_{m_2}[K_{m_2}A]=L_{m_1}[K_{m_2}A]
\]
This means that the vacuum functionals can be interchanged {\it after} the
mapping but not before.

\item [5.] The kernels ${\cal K}_m$ determine the representation of the
Poincar\'e group in the light-front mass m sub-algebra.  Specifically
the infinitesimal versions of the covariant Poincar\'e transformations,
$x^{\mu} \to x^{\prime\mu} \to \Lambda^{\mu}{}_{\nu}+ a^{\mu}$, get mapped
into the light-front generators defined as operators on the
light-front Fock algebra.  The kernels also satisfy the field equations.

\item [6.] The kernels, ${\cal K}_m$, for different masses are associated with
  inequivalent representations of the canonical commutation relations.
  This follows because vacuum expectation values of
  polynomials in the Heisenberg fields can be expressed in terms of
  vacuum expectation values of both polynomials in
  the canonical algebra or vacuum expectation values of
  polynomials in the light-front mass $m$ sub-algebra.

\item [7.] The kernels ${\cal K}_m$ determine local nature of the
Heisenberg fields.  This follow directly from the operator relation
(\ref{ir22}).  Specifically is $f$ and $g$ have space-like separated
supports then
\[
0=  [\phi(f),\phi(g)]_{\pm}= [\phi_{LF}(f_m),\phi_{LF}(g_m)]_{\pm}
\]
where the $+$ is for half integer spin free fields and the $-$ is for integer
spin free fields.

\item [8.] The kernels map Schwartz functions to functions of light-front variables that vanish faster than any polynomial in $p^+$ near
  $0$. It will be shown that this suppression of light-like momenta in
  the light-front mass $m$ test functions removes non-causal momenta
  from the light-front mass $m$ algebra resulting in a well-defined
  initial value problem.

\end{itemize}

Since the light front fields are still operator valued distributions the
fields on the light front must first be smeared with
test functions before taking vacuum expectation values.   In order to calculate the restriction of the Heisenberg field to the light front it is necessary to take the limit using the
kernel ${\cal K}_m$.  For free fields the result will be
\cite{bogoliubov:1959}
\beq
\langle 0 \vert \phi (x) \phi (0) \vert 0 \rangle  \to 
-i {\epsilon (x^-) \delta (\mathbf{x}_{\perp}^2) \over 4 \pi} 
- { m  K_1 (m \sqrt{\mathbf{x}^2_{\perp}} )
\over 4 \pi^2 \sqrt{\mathbf{x}_{\perp}^2 }}
.
\label{ir25}
\eeq
On the other hand if this is computed directly on the light front without using one of the kernels ${\cal K}_m$, then the result will be
\beq
\langle 0 \vert \phi (x) \phi (0) \vert 0 \rangle =
{1 \over 2 (2 \pi)^3} \int {\theta (q^+)dq^+ d\mathbf{q}_{\perp} \over q^+ }
e^{-i {q^+\over 2}x^-
  + i \mathbf{q}_{\perp} \cdot \mathbf{x}_{\perp}} =
{\delta (\mathbf{x}_{\perp}^2) \over 4 \pi}
\int_0^{\infty} {dq^+ \over q^+ }
e^{-i {q^+\over 2}x^-}
\label{ir26}
\eeq
again independent of the choice of vacuum.

\section{Initial value problem}

A dense set of Schwartz functions are Fourier transforms of Schwartz
functions with compact support. These
functions satisfy
\beq
\{f(x) \vert \hat{f} (p) := \int {d^4x \over (2 \pi)^2}e^{-p\cdot x}f(x), 
\qquad \hat{f} (p) = 0 \qquad \mbox{for} \qquad  (p^0)^2 + \mathbf{p}^2 < R^2 < \infty\}
\label{iv1}
\eeq
where $R$ is any finite positive constant.  The Heisenberg algebra of fields
smeared with these test functions applied to the vacuum generates a dense subspace of the Hilbert space.

If $\hat{f}(p)$ is expressed in terms of light front variables
$\hat{f}(p) = \hat{f}({p^++p^- \over 2} ,\mathbf{p},{p^+-p^- \over 2} )
=: g(p^-,p^+,\mathbf{p}_{\perp})$, for $g(p)$ to be non-vanishing
requires
\beq
\vert p^+ \vert, \vert p^- \vert < 2R.
\label{iv2}
\eeq
The Fourier transform of the light front mass $m$ test function
\beq
g_m(\tilde{\mathbf{y}}) =\int {\cal K}_m (x,\tilde{\mathbf{y}}) f(x) d^4 x
\label{iv3}
\eeq
is (\ref{ir23}) 
\beq
\hat{g}_m (p^+,\mathbf{p}_{\perp}) =
\sqrt{2 \pi} \hat{g}({\mathbf{p}^2+m^2 \over p^+},p^+, \mathbf{p}_{\perp}).
\label{iv4}
\eeq
The support condition implies 
\beq
\hat{g}(p) =0  \qquad \mbox{unless} \qquad (p^0)^2 + \mathbf{p}^2 < R^2 < \infty
\label{iv5}
\eeq
for some finite $R$.  The support condition (\ref{iv2}) implies that
\beq
\vert p^+ \vert <2R \qquad \mbox{and} \qquad \vert p^-\vert ={\mathbf{p}^2+m^2 \over p^+} <2R .
\eeq
These inequalities can be combined to show that the support of 
$\hat{g}_m (p^+,\mathbf{p}_{\perp})$ is for 
\beq
{m^2\over 2R } <p^+ < 2R .
\label{iv6}
\eeq
This condition eliminates light-like momenta with
$p^+=0$ provided $m>0$.

The $x^+$ evolution in the covariant representation is a translation of the
argument of the test function.
The intertwining relations (\ref{ir24}) give 
\[
\int \phi (x^+, \tilde{\mathbf{x}}) f(x-a^+)d^4 x =
\]
\[
\int \phi (y^++a^+, \tilde{\mathbf{y}}) f(y^+,\tilde{\mathbf{y}})d^4 y =
\]
\[
\int d^4y d\tilde{\mathbf{p}} 
f(y)
{\cal K}_m (y^+ + a^+,\tilde{\mathbf{y}} ,\tilde{\mathbf{p}})\phi_{LF}
(-\tilde{\mathbf{p}}) =
\]
\beq
\int d^4y d\tilde{\mathbf{p}} 
f(y)
\sum_{n=0}^\infty {(ia^+)^n \over n!}
{\partial^n \over \partial ({iy^+})^n} 
{\cal K}_m (y^+,\tilde{\mathbf{y}} ,\tilde{\mathbf{p}})
\phi_{LF}  (-\tilde{\mathbf{p}}) .
\label{iv7}
\eeq
Using (\ref{ir24}) gives
\[
=\int d^4y d\tilde{\mathbf{p}} 
f(y)
\sum_{n=0}^\infty {(-i a^+)^n \over n!}
({\mathbf{p}^2 + m^s \over 2p^+})^n
{\cal K}_m (y^+,\tilde{\mathbf{y}} ,\tilde{\mathbf{p}})
\phi_{LF}  (-\tilde{\mathbf{p}})= 
\]
\[
\int d\tilde{\mathbf{p}}
e^{-i a^+({\mathbf{p}^2 + m^s \over 2p^+})}
f_m (\tilde{\mathbf{p}})   
\phi_{LF} (-\tilde{\mathbf{p}}) =
\]
\beq
\sqrt{2 \pi} \int d\tilde{\mathbf{p}}
\sum_{n=0}^\infty {(-i a^+)^n \over n!}
({\mathbf{p}^2 + m^s \over 2p^+})^n
\hat{f} ({\mathbf{p}_{\perp}^2+m^2 \over p^+},\tilde{\mathbf{p}})
\phi_{LF} (-\tilde{\mathbf{p}}) .
\label{iv8}
\eeq
The series
\beq
\sum_{n=0}^\infty {(-ia^+)^n \over n!}({\mathbf{p}_{\perp}^2+m^2 \over 2 p^+})^n
\hat{f}({\mathbf{p}_{\perp}^2+m^2 \over p^+},\tilde{\mathbf{p}})
\label{iv9}
\eeq
converges provided for $m>0$ since
\beq
\vert \sum_{n=0}^\infty {(-ia^+)^n \over n!}({\mathbf{p}_{\perp}^2+m^2 \over 2 p^+})^n\vert \leq \sum_{n=0}^\infty {(2Ra^+)^n \over n!} =e^{2Ra^+} < \infty .
\label{iv10}
\eeq
This shows that the initial value problem for the field is well
defined on the light front provided the mass m test functions are
derived from this dense set of Schwartz functions.

\bigskip

The relevant observations are:
\begin{itemize}
\item[1] The mapping from the Heisenberg field algebra to the light-front mass $m$ field algebra maps the covariant representation of the Poincar\'e group to a light-front representation (\ref{ir24}).

\item[2.] This mapping maps a dense set of vectors in the covariant representation of the Hilbert space to a dense set of vectors in the space generated by
the light-front mass m sub-algebra.

\item[3.] $x^+$ evolution, defined by the exponential series, converges
on this dense set of vectors.

\item[4.] The dense set in the Hilbert space can be generated by applying
elements of this  sub-algebra of the light-front Fock algebra to any
free-field vacuum.

\end{itemize}

\section{Triviality of the vacuum}

Since the vacuum is invariant with respect to translations
that leave the light-front invariant, it follows that
\beq
P^+_0 \vert 0 \rangle =0 .
\label{tv1}
\eeq
Since $P^+_0$ commutes with both $M$ and $M_0$ it
commutes with the interaction
$V:=M-M_0$.  It follows that
\beq
P_0^+ V \vert 0 \rangle = V P_0^+ \vert 0 \rangle =0 .
\label{tv2}
\eeq
This means that $\vert 0 \rangle$ and $V\vert 0 \rangle$ are both eigenstates
of $P^+_0$ with eigenvalue $0$.  It follows that
\beq
\langle 0 \vert V^{\dagger}V \vert 0 \rangle =
\int \vert \langle p_0^+ , d \vert V \vert 0 \rangle\vert^2
d\mu (p_0^+)  dd 
\label{tv3}
\eeq
where $d$ represents the remaining degenerate quantum numbers and
$d\mu (p_0^+)$ is the spectral measure of $P_0^+$.  The spectrum of
$P^+_0$, except for the vacuum, is absolutely continuous and
non-negative.  The only contribution to this integral over
intermediate states comes from states that are discrete eigenstates of
$p_0^+$ with eigenvalue $0$.  It follows that
\beq
V\vert 0 \rangle = \vert 0 \rangle \langle 0 \vert V \vert 0 \rangle
\label{tv14}
\eeq
which means that $V$ cannot change the vacuum.

If the vacuum is the only discrete normalizable
state of the theory that is invariant under translations on the
light front then
\[
0 = (P^-P^+ -\mathbf{P}_{\perp}^2) \vert 0 \rangle  = M^2 \vert 0 \rangle =
(M_0^2+VM_0 + M_0V + V^2 )\vert 0 \rangle =
\]
\beq
V^2 \vert 0 \rangle =
\vert 0 \rangle \langle 0 \vert V \vert 0\rangle^2
\label{tv5}
\eeq
which means that the constant must vanish.

The problem with this naive argument is that all of the
translation generators are unbounded operators.  $P^+=0$ corresponds
to infinite momentum while $P^-$ diverges as $p^+ \to 0$, so
$P^+\vert 0 \rangle=0$ does not imply $P^- P^+\vert 0 \rangle=0$.
This is clear since the light-front dispersion relation has the form
$P^-= (M^2 +\mathbf{P}_{\perp}^2)/P^+$, which has a $P^+$ in the denominator.

In a dynamical theory
the pure creation operator part of the interaction
is the part of the interaction that is responsible for
the difference between the Heisenberg and Fock vacuums
in canonical formulations of quantum field theory.
In a $\phi^4(x)$ theory the pure creation part of the interaction term
in the light-front Hamiltonian has the form:
\bigskip
\[
\int {\theta (p^+) \delta (p^+) dp^+\over (p^+)^2 \prod \xi_i}
\prod d\mathbf{p}_{i\perp} d\xi_i \delta (\sum \mathbf{p}_{i\perp})
\delta (\sum \xi_i -1) \times
\]
\[
a^{\dagger}(\xi_1 p^+, \mathbf{p}_{\perp 1})
a^{\dagger}(\xi_2 p^+, \mathbf{p}_{\perp 2})
a^{\dagger}(\xi_3 p^+, \mathbf{p}_{\perp 3})
a^{\dagger}(\xi_4 p^+, \mathbf{p}_{\perp 4}). 
\]
This expression is divergent as $p^+\to0$.  Multiplication by
$p^+$ is not sufficient to cancel the $p^+=0$ singularities in the
denominator.

It is clear some kind renormalization is necessary to
remove this singular behavior.  In this next section is
will be argued that how this must be done is constrained
by rotational covariance.  The interacting case
will be discussed in the following section.

\section{Rotational Covariance}

The light front Hamiltonian $P^-$ and the kinematic generators of the
Poincar\'e group form a closed sub-algebra.  This sub-algebra contains no
information about transverse rotations.  Given this sub-algebra, if
${J}^1$ and ${J}^2$ are the transverse rotation
generators that complete the Poincar\'e Lie algebra and $W$ is any
unitary operator that commutes with $P^-$ and the kinematic generators
then ${J}^{\prime 1}= W{J}^{1}W^{\dagger}$ and
${J}^{\prime 2}= W{J}^{2}W^{\dagger}$ also complete the
Poincar\'e Lie algebra with the same $P^-$ and kinematic generators.
It follows that $P^-$ and the kinematic generators do not uniquely
determine the transverse rotation operators.

Noether's theorem, which assumes the equations of motion, provides
conserved currents for all ten independent Poincar\'e transformations
(see Appendix II).  Noether charges can be constructed by integrating
the current over a light front.  Even though the currents are
integrated over a light front rather than a fixed time surface, the
resulting Noether charges for free fields satisfy the Poincar\'e
commutations relations.  In the light-front case both the kinematic
and dynamical generators are expressed in terms of the irreducible
algebra of fields restricted to the light front.  There are explicit
expressions for both $P^-$ and the transverse rotation generators,
$\mathbf{J}_{\perp}$.
So while $P^-$ and the kinematic generators are not sufficient to complete the
Poincar\'e Lie algebra,  Noether's theorem gives explicit
expressions for the transverse rotation generators in terms of fields
on the light front that correspond to a particular choice of $W$.

The transverse rotation generators do not have this ambiguity. 
The Poincar\'e commutation relations imply 
\beq 
P^-:= P^+ -2 [J^2,[J^2,P^+]] \qquad \mbox{and} \qquad J^1:=-i [J^2,J^3]
\label{RI1}
\eeq
which means that given $J^1$ or $J^2$ and the kinematic generators it
is possible to use the commutation relations to construct the other
two dynamical generators.  Since these relations are a consequence of the
Poincar\'e commutation relations they also hold in dynamical theories.

The rotational properties of fields can be
solved using the commutation relations. 
Formally rotations about the $y$ axis can
be expressed as an infinite  series in the irreducible light front algebra
\[
U_y(\theta) \phi(\tilde{\mathbf{x}},0) U_y^{\dagger}(\theta) =
\phi (
     {1 + \cos (\theta) \over 2}x^-,- \sin(\theta) x_1,x_2, {1 - \cos (\theta) \over 2}x^-)=
     \]
     \beq
\sum_{n=0}^\infty {(i \theta)^n \over n!}
\underbrace{[J_2,[J_2,\cdots , [J_2, \phi (\mathbf{\tilde{x}},0)]]]
}_{\mbox{N times }}.
\label{RI2}
\eeq
where $J^2$, $\phi (\mathbf{\tilde{x}},0)$ and all commutators are all restricted to
the light front.  The arguments of the coordinates $x^-$ and
$\mathbf{x}_{\perp}$ on the light front can be changed to any values
using kinematic translations.  Equation (\ref{RI2}) uses the fields on
the light front and the rotation generators, constructed from
fields on the light front, to formally construct the
local Heisenberg field algebra.

The explicit structure for $J^2$ (for scalar fields) in terms of field operators
restricted to the light front that follows from Noether's theorem is:
\[
J^2 = -\int_{x^+=0} {d\mathbf{x}_{\perp} d x^- \over 4} 
\left(
x^- T^{+1}(x) + x^1 (T^{++}- T^{+-})) \right ) =
\]
\beq
\int_{x^+=0} {d\mathbf{x}_{\perp} d x^- \over 4} 
\left(
- x^- 2 :\partial_- \phi(x) \partial_1 \phi(x):
\right .
\left .
- x^1 (4 :\partial_- \phi(x) \partial_- \phi(x): -  (
 :\pmb{\nabla}_{\perp} \phi(x) \cdot
\pmb{\nabla}_{\perp} \phi(x): + {m^2} :\phi(x)^2: + 2:V(\phi(x)):
) \right )
\label{RI3}
\eeq
where $T^{\mu\nu}$ is the energy momentum tensor density.   The relevant observation is that 
this expression only involves fields on the light front and derivatives
of fields that are tangent to the light front.

Equation (\ref{RI2}) is formally true for both free and interacting
theories, but in the interacting case the operator products in
$:V(\phi(x)):$ have divergences as $p^+\to 0$ and $p \to \infty$ that
require renormalization in order to define $J^2$.

While it is formally sufficient to construct the dynamics using $J^2$
and kinematic transformations following the construction in the free
field case, the structure of the dynamical part of $J^2$ is
constrained by the Poincar\'e commutation relations.  While Noether's
theorem gives the formal expression (\ref{RI3}) for $J^2$, the
complete definition requires renormalization of the operator products
which must be done in a manner that preserves these constraints.

To understand the origin of these constraints it is useful to express
the transverse rotation generator, $J^2$,  as the sum of a
non-interacting generator and an interaction, $J^2= J^2_0+ J^2_I$,
where $J_0^2$ is the kinematic generator of rotations about the 2
axis. The commutation relations imply that the interaction, $J^2_I$,
must commute with all of the kinematic generators and in addition it must
satisfy the following linear and non-linear dynamical constraints:
\beq
[J_I^2 , [J_0^2,P^1]] =0
\label{RI4}
\eeq
and
\beq
[J_I^2 , [J_I^2,J^3]]+[J_0^2 , [J_I^2,J^3]]
+i [J_I^2 , J_0^1] =0.
\label{RI5}
\eeq
These are non-trivial constraints on the allowed interactions in $J^2$. 

In the absence of the other dynamical generators the renormalization
of $P^-$ is not constrained by rotational covariance.  This is the
situation that appears in many applications of light-front field
theory.  The rotational covariance is not encoded in $P^-$ so it must
be imposed by hand.  For free fields the angular momentum is already
encoded in the spinors for the field restricted to the light front.
In an interacting theory there can be degenerate
composite eigenstates of mass and $\mathbf{S}\cdot \hat{\mathbf{z}}$
where $P^-$ provides no information on how to assign the total spin to
these states.  The transverse angular momentum operator is needed to
separate these states according to their spin.  Knowing only $P^-$ is
not sufficient.

A necessary condition for light-front formulations of quantum field theories
to be equivalent to the covariant formulation is that quantum
observables do not depend on the choice of orientation of the light
front \cite{Karmanov1}
\cite{Karmanov2}
\cite{Karmanov3}
\cite{Karmanov4}
\cite{Karmanov5}
\cite{Fuda:1994uv}\cite{Fuda:1990}
\cite{Polyzou:1999}.  Let $U_{\hat{\mathbf{n}}} (g)$, with $g=(\Lambda,a)$ be the
unitary representation of the Poincar\'e group associated with light
front $x^0+\hat{\mathbf{n}}\cdot \mathbf{x}=0$ and let
$K_{\hat{\mathbf{n}}}$ denote the corresponding kinematic
subgroup. The operator
\beq
Y_{\hat{\mathbf{n}}}(g) := U_0(g) U^{\dagger}_{\hat{\mathbf{n}}}(g)
\label{RI6}
\eeq
is unitary and is the identity whenever $g\in K_{\hat{\mathbf{n}}} $.
This means that $Y_{\hat{\mathbf{n}}}(g)$ only depends on
left cosets of $K_{\hat{\mathbf{n}}}$ in the Poincare group:
\beq
g' \in [g]_{\hat{\mathbf{n}}} \iff g' = gk \qquad \mbox{for some} \qquad
k \in K_{\hat{\mathbf{n}}}. 
\label{RI7}
\eeq
It follows that (\ref{RI6}) can be expressed as 
$
Y_{\hat{\mathbf{n}}}([g]_{\hat{\mathbf{n}}}).
$
Applying this operator to $U_{\hat{\mathbf{n}}} (g)$ gives a new unitary representation of the
Poincar\'e group with a different kinematic subgroup
\beq
Y_{\hat{\mathbf{n}}}([g]_{\hat{\mathbf{n}}})  U_{\hat{\mathbf{n}}} (g)
Y^{\dagger}_{\hat{\mathbf{n}}}([g]_{\hat{\mathbf{n}}})
=
U_{\hat{\mathbf{n}}'} (g)
\label{RI9}
\eeq
where the new kinematic subgroup is 
\beq
K_{\hat{\mathbf{n}}'} = [g] K_{\hat{\mathbf{n}}} [g^{-1}] .
\label{RI10}
\eeq
If in addition this transformation leaves the scattering operator
unchanged it can be expressed as in terms of wave operators
\beq
Y^{\dagger}_{\hat{\mathbf{n}}}([g]_{\hat{\mathbf{n}}}) =
\Omega_{+K}\Omega^{\dagger}_{+[g]K[g^{-1}]} =
\Omega_{-K}\Omega^{\dagger}_{-[g]K[g^{-1}]} 
\label{RI11}
\eeq
which means that the operator can be expressed in terms of complete
sets of scattering solutions with different light fronts.  Using these relations with (\ref{RI6}) for $g=(R,0)$, a pure rotation, gives an explicit expression of the rotation operator with kinematic subgroup $K_{\hat{\mathbf{z}}}$:   
\beq
U_{\hat{\mathbf{z}}}(R,0)= Y^{\dagger}_{\hat{\mathbf{z}}}([R]_{\hat{\mathbf{z}}}) U_0(R)
 =
\Omega_{\pm K_{\hat{\mathbf{z}}} }\Omega^{\dagger}_{\pm [R]K_{\hat{\mathbf{z}}}[R^{-1}]} U_0(R).
\label{RI12}
\eeq
This provides an expression for the rotation operator in terms of a complete sets of wave
functions for the same state in different S-matrix equivalent
representations of the dynamics.  These formal relations hide the
actual difficulty.  In order to calculate the scattering states it is
important to know the rotational properties of the asymptotic states.
If there are 2 composite one-body solutions with
the same mass that are degenerate in the magnetic quantum numbers, it
is necessary to know how to consistently assign total spins the these
states in the asymptotic region order to assign them to the
appropriate scattering channel.  Equation (\ref{RI12}) assumes that
problem has been solved, but it is equivalent to the problem of
assigning spins to degenerate bound states in light-front dynamics.

If this is done successfully it gives a dynamical representation of the
rotation group.  While it is not necessarily that same as the one
that comes from the Noether charges, it will be unitarily equivalent.

The important observation about the relation between rotational
covariance and invariance with respect to changing the orientation of
the light front is that if theories with different light front
orientations are scattering equivalent, the $p^+\to 0$ divergences 
associated with one light front become $p\to \infty$ divergences when
associated with a different light front.  This means that the $p^+\to 0$ 
and $p\to \infty$  renormalizations are constrained by rotational
covariance.  At the perturbative level the connection can be resolved
by appealing to the connection to covariant perturbation theory
\cite{Mannheim:2020rod}, but one of the compelling reasons for using
the Hamiltonian formulation of light-front quantum field theory is that
it is in principle a non-perturbative formulation of quantum field theory.
This means the problem of rotational covariance cannot be separated from the
problem of how to perform rotationally invariant renormalization
beyond perturbation theory.

\section{Interacting Fields}

For case of interacting fields, in the absence of non-trivial
dynamical solutions, light-front field theories can be studied based
on general properties of quantum field theory.  This assumes that
the problem of how to properly renormalize the theory outside of
perturbation theory has been solved. 

The first assumption is that the theory is asymptotically complete.
This means that it has complete sets of scattering states satisfying
IN and OUT asymptotic conditions and these states (along with the
vacuum and one-particle states) are a basis for the Hilbert space of
the theory.  The convergence of the limits that define these
scattering states is a consequence of the axioms of quantum field
theory provided the theory has one-particle states
(point spectrum eigenstates) and a mass gap.
The scattering states are expressed as strong limits in the Haag-Ruelle
formulation of scattering theory
\cite{jost:1966}\cite{simon}\cite{araki-bk}\cite{strocchi}.

The Haag-Ruelle formulation of scattering is a field theoretic version of the
non-relativistic formulation of time-dependent scattering 
where the scattering states are defined by the strong limits
\beq
\vert f_1 \cdots f_n \rangle_{OUT/IN} =
\lim_{t \to \pm \infty}
\int 
\phi_h(\mathbf{x}_1,t)\cdots \phi_n(\mathbf{x}_N,t) \vert 0 \rangle
\stackrel{\leftrightarrow}{\partial}_t
\prod_n f_n (\mathbf{x}_n,t ) d \mathbf{x}_n ,
\label{HR1}
\eeq
and the functions,
\beq
f_n (\mathbf{x}_n,t ) =
\int  
{e^{-i \omega_n(\mathbf{p})t +i \mathbf{p}\cdot \mathbf{x}}
\over (2 \pi)^{3/2}(2\sqrt{\omega_m(\mathbf{p})})} g (\mathbf{p})
d\mathbf{p} 
\label{HR2}
\eeq
are positive energy solutions of the Klein Gordon equation. 
$\phi_h(x)$ is a quasilocal field which is the Fourier transform of
\beq
\hat{\phi}_h(p) = \hat{\phi}(p) h(p)  
\label{HR3}
\eeq
where $\hat{\phi}(p)$ is the Fourier transform of the field
and $h$ is
a smooth function that is 1 when $p^2=-m^2$ and $p^0>0$ and vanishes on
the rest of the mass spectrum of the theory and for $p^0<0$.  Here
the mass is the physical mass of the one-particle state.  If the
theory has one-particle states that are not associated with the elementary
fields $\hat{\phi}(p)$ can be replaced by the Fourier transform of
a local interpolating field \cite{Zimmermann:1958} that has non-zero matrix elements between the vacuum and one-particle state .

Ruelle showed that the strong limit of (\ref{HR1}) exists when
$g(\mathbf{p})$ is a smooth function of the 3-momentum with compact
support.  The integral of the norm of the difference of two sides of
(\ref{HR1}) has the same kind of $t^{-3/2}$ fall off that occurs in
the Cook condition in quantum mechanical scattering, except in this
case the fall off is due to cluster properties and the assumed mass
gap rather than the range of the potential.

The purpose of providing this detail is the observation that the
scattering states can be expressed as an operator applied to the
vacuum that is multi-linear in the wave packets $g_i(\mathbf{p})$ which
correspond to the momentum distributions of the asymptotic clusters in
the scattering reaction.  It is also multi-linear in the positive
energy solutions $f(\mathbf{x},t)$ of the Klein Gordon equation.  The
time derivative, $\stackrel{\leftrightarrow}{\partial}_t$, selects the
part of the field that asymptotically looks like a creation operator.

The set of $IN$ or $OUT$ scattering states form a complete orthonormal
set of states in the theory.  These strong limits can be used to
define asymptotic field operators 
that transform N-particle scattering states to $N\pm 1$ particle
scattering states.  
\beq
\Phi_{IN} (f) \vert f_1 \cdots f_n \rangle_{IN} =
\vert f,f_1 \cdots f_n \rangle_{IN}
\label{HR4}
\eeq
where 
\beq
\Phi_{IN} (f) = \lim_{t \to -\infty} \int \Phi_{IN} (\mathbf{x},t)  \stackrel{\leftrightarrow}{\partial} f(\mathbf{x},t) d\mathbf{x}.
\label{HR5}
\eeq
There are IN (resp. OUT) fields for each one-particle state of the
field theory.  Mathematically $\Phi_{IN}$ has the same Fourier
representation as free fields with physical masses.  The difference is
that the creation operators create $n+1$ particle scattering states
out of $n$ particle scattering states.  The algebra of IN (resp OUT)
fields applied to the vacuum generates the Hilbert space of the field
theory.  Mathematically it is equivalent to a free field Fock space
(of a direct sum of free field Fock spaces).
Just like ordinary free fields the Fourier transforms of the
asymptotic fields can be expressed in terms of light-front variables,
and as in the case of free fields the creation and annihilation
operators in the asymptotic fields can be expressed in terms of the algebra
of asymptotic fields
restricted to the light front.  This follows from the construction in
section V:
\beq
\int \Phi_{IN} (x)f^*(x) d^4 x=
\int {\cal K}_m (x,\tilde{\mathbf{y}}) f^*(x) d^4x =
\int \Phi_{IN} (y^+=0,\tilde{\mathbf{y}})f_m (\tilde{\mathbf{y}})
{dy^-d\mathbf{y}_{\perp}\over 2}.
\label{HR6}
\eeq
This generates a map from the algebra of asymptotic fields
to a sub-algebra of the light front field algebra as in the case of free fields.

The assumed irreducibility of the asymptotic fields means that the
Heisenberg fields can be expanded in normal products of the asymptotic
fields:
\cite{Haag:1958vt}\cite{Glaser:1957}\cite{Greenberg:1965}\cite{yangfeldman}:
\beq
\Phi(f) =
\sum \int R_n(x;x_1 \cdots x_n)f(x) d^4x
:\Phi_{IN} (x_1) \cdots \Phi_{IN} (x_n) : \prod_k d^4x_k .
\label{HR7}
\eeq
Using (\ref{HR6}) this expansion
can be expressed in terms of the algebra of fields on the light front
\beq
\Phi(f) =
\sum \int R_n(x;x_1, \cdots ,x_n)f(x) d^4x
\prod_k d^4x_k K_{m_k} (x_k,\tilde{\mathbf{y}}_k) 
:\Phi_{IN} (y_1^+=0, \tilde{\mathbf{y}}_1)\cdots
\Phi_{IN} (y_n^+=0, \tilde{\mathbf{y}}_n):
\prod_l d\tilde{\mathbf{y}}_l .
\label{HR7}
\eeq
This expansion defines the analog of the mapping (\ref{ir21}) from the
Heisenberg field algebra to a sub-algebra of the light-front
Fock algebra.   Some versions of perturbation theory start with
expansions Heisenberg fields in terms of asymptotic fields rather than free fields
\cite{Veltman}.

Vacuum expectation values of products of the Heisenberg fields can be expressed
as light-front expectation values of functions of the asymptotic fields
restricted to the light front using the expansions (\ref{HR7}). 
As in the case of free fields, any vacuum restricted to the light
front can be used in this computations.  The mapping defines the
values of the vacuum functional off of the light-front.

If there are asymptotic states with composite particles
there will be
tensor products of Fock spaces of asymptotic states restricted to
the light-front.  Since the Fock space are countable they are
unitarily equivalent to tensor products of Fock spaces, so
vacuum expectation values of products of smeared asymptotic field can still
be expressed in terms of vacuum expectation values of free fields
on the light-front.

One of the primary purposes of using the Hamiltonian formulation of
light-front quantum field theory is applications to QCD.  One of the
new problems is that the asymptotic fields for QCD are all composite
fields.  Individual Heisenberg fields for quarks and gluons cannot be
represented as expansions in the asymptotic fields of the theory,
which are all colorless.  Greenberg \cite{Greenberg:1978} showed that
in a simple model the introduction of ``confined asymptotic fields''
satisfying certain properties allowed for the expansion of color
carrying field in terms of normal products of asymptotic fields.
Whether this is applicable to QCD remains an open question.  This does
not imply that the irreducible algebra of fields on the light-front
cannot be used to represent smeared Heisenberg quark fields.  The
light front is particularly interesting because the dependence on the
quark masses disappears for fields restricted to the light front.
This consistent with the partonic description of quarks, which have
electric charge but no asymptotic mass.

The main observation is that even in an interacting theory, if the
theory is properly defined,  the Wightman functions of the theory can
be expressed in terms of a sub-algebra of fields on the light front
applied to any vacuum restricted to the light front.  In this
case the sub-algebra distinguishes different theories.

This assumes that the full mass spectrum has been determined.  In order to
use this it is necessary to determine the point spectrum of the theory
and express the Heisenberg fields in terms 
of normal products of the IN or OUT fields.

Just like in the free field case,  once the smeared Heisenberg fields
are expressed in as elements of a sub-algebra of the light front algebra,
vacuum expectation values of products of fields can be evaluated using any
vacuum.  In this case the vacuum looks trivial because the true vacuum
is equivalent to a free field vacuum on the light front sub-algebra
associated with the interacting theory.

\section{Spontaneous Symmetry Breaking}

One of the puzzling aspects of light-front quantum field theory is how
spontaneous symmetry breaking is realized.  The signal for spontaneous
symmetry breaking is a conserved local current, $j^{\mu}(x)$, that
arises from the symmetry and the presence of a 0 mass Goldstone boson
in the spectrum.  The presence of a Goldstone boson is confirmed when
charge operator has non-zero matrix elements between the vacuum and
Goldstone boson.  In general the charge operator does not have to make
sense, since it is an operator valued distribution evaluated at point
$\mathbf{p}=0,t=0$.

For a local scalar field theory the following condition \cite{coleman}
\beq
\lim_{R \to \infty} \langle 0 \vert [Q_R, \phi(y)] \vert 0 \rangle \not=0,
\label{SB:1}
\eeq
where 
\beq
\langle 0 \vert [Q_R, \phi(y)] \vert 0 \rangle :=
\langle 0 \vert [ \int d\mathbf{x} \chi_R(\vert \mathbf{x} \vert)
j^0(\mathbf{x},t), \phi(y)] \vert 0 \rangle
\label{SB:2}
\eeq
where $\chi_R(\vert \mathbf{x} \vert)$ is a smooth function that
that is 1 for $\vert \mathbf{x}\vert <R$ and $0$ for $\vert \mathbf{x}\vert
>R+\epsilon$ for some finite positive $\epsilon$,
which does not require the existence of a charge operator, 
implies the existence of a 0 mass particle.

The cutoff function $\chi_R$ ensures that the integral converges for 
large $\vert\mathbf{x}\vert$.  The commutator vanishes for
$x-y$ space-like.  For fixed $y$ and $t$ and sufficiently large
$R$, if $\vert \mathbf{x}\vert >R$ then $x-y$ is space-like and
the commutator vanishes.  In this case
\beq
\langle 0 \vert [Q_R, \phi(y)] \vert 0 \rangle :=
\langle 0 \vert [ \int d\mathbf{x} \chi_R(\vert \mathbf{x} \vert)
j^0(\mathbf{x},t), \phi(y)] \vert 0 \rangle =
\int d\mathbf{x} \langle 0 \vert [ 
j^0(\mathbf{x},t), \phi(y)] \vert 0 \rangle .
\label{SB:3}
\eeq

Current conservation implies
\beq
\langle 0 \vert [ \int d\mathbf{x}
\partial_{\mu} j^\mu(\mathbf{x},t)
, \phi(y)] \vert 0 \rangle = 0 .
\label{SB:4}
\eeq
Inserting a complete set of intermediate states gives
\[
0=\sum \int d\mathbf{x} \left(
\langle 0 \vert \partial_{\mu} j^\mu(\mathbf{x},t)
\vert p ,n \rangle 
{d\mathbf{p} \over 2p_n^0}  \langle p,n \vert \phi(y) \vert 0 \rangle
- \langle 0 \vert  \phi(y)
\vert p ,n \rangle {d\mathbf{p} \over 2p_n^0}
\langle p,n \vert \partial_{\mu}
j^\mu(\mathbf{x},t)  \vert 0 \rangle\right )=
\]
\beq
\sum\int d\mathbf{x} \left(
\langle 0 \vert \partial_{\mu} j^\mu(\mathbf{0 },0)
\vert p_r ,n \rangle {d\mathbf{p} \over 2p_n^0} \langle p_r,n \vert \phi(0) \vert 0 \rangle
e^{i p\cdot (x-y)}
- \langle 0 \vert  \phi(0)
\vert p_r ,n \rangle {d\mathbf{p} \over 2p_n^0} \langle p_r,n \vert \partial_{\mu}
j^\mu(\mathbf{0},0)  \vert 0 \rangle
e^{i p\cdot (y-x)}
\right )
\label{SB:5}
\eeq
where Poincar\'e covariance has been used to remove the non-trivial
$x,y$ and $p$ dependence from the matrix elements. $p_r$ is
the constant rest four momentum for massive states and a constant light-like
vector for massless states.  For a scalar field theory the vacuum
expectation value of the current vanishes so the vacuum does
not appear as an intermediate state.  The Lehmann weights that appear in
this matrix element
\beq
\sigma (m_n)m_n^2 = 
\langle 0 \vert \partial_{\mu} j^\mu(\mathbf{0},0)
\vert p_{nr} \rangle \langle p_{nr} \vert \phi(0) \vert 0 \rangle
\label{SB:6}
\eeq
and
\beq
\sigma^*(m_n)m_n^2 =
\langle 0 \vert  \phi(0)
\vert p_{nr} ,n \rangle \langle p_{nr},n \vert \partial_{\mu}
j^\mu(\mathbf{0},0)  \vert 0 \rangle
\label{SB:7}
\eeq
are functions of the invariant mass of the intermediate states.
The factor $m_n^2=-p_n^2$ is due to the $\partial_{\mu} j^\mu(\mathbf{0},0)$
assuming that the current is a local function of the scalar field.
Current conservation requires that this quantity vanishes which
follows if either $\sigma(m_n)=0$ or $m_n^2=0$.
Inserting intermediate states in (\ref{SB:2}) gives a similar result
with $\sigma (m_n)m_n^2$ replaced by  
$p_{nr}^0\sigma (m_n)$ where $p_{nr}^0$ is a non-zero constant.

Since this does not include the $m^2$ factor,
it vanishes by current conservation unless $\sigma(m_n)$
contains a $\delta (m_n)$.  Thus for (\ref{SB:1}) to be non-vanishing 
the sum over intermediate states must have a contribution from
a 0 mass particle.   This is the non-perturbative
form of Goldstone's theorem.

In \cite{coleman} Coleman avoids directly computing the charge
operator.  The current is an operator valued distribution, so there is
no reason to expect that integrating against the constant $1$ will
result in a well defined operator.

While the non-vanishing of (\ref{SB:1}) is a sign of spontaneous
symmetry breaking, the argument used above does not work when it is
applied to the light-front charge.  The problem is that there is no
compact region on the light front where outside of that region $(x-y)$
is always space-like for fixed $y$ and $x^+=0$.  In the light-front case, if the
charge exists, $Q$ is invariant with respect to translations on the light front,
which means that it commutes with $P^+$.  It follows that  
$P^+ Q \vert 0 \rangle= Q P^+ \vert 0 \rangle =0 $.  Because of the spectral
condition all of the individual $p_i^+$,  $p_i^+ Q \vert 0 \rangle =0$.
Multiplying $Q \vert 0 \rangle$ by the completeness relation implies that
the only non-zero term in the sum is the vacuum
\beq
Q \vert 0 \rangle = \vert 0 \rangle \langle 0 \vert  Q \vert 0 \rangle
\label{SB:8}
\eeq
which means that it cannot couple the vacuum to a another state.

In the light-front representation it is still possible to define
$Q_R$ on a fixed-time surface and the analysis above still applies.
To do this the field and current in the matrix element (\ref{SB:1})(\ref{SB:2})
must
be mapped to the sub-algebra of the light-front Fock algebra.
Since spontaneous symmetry breaking is a dynamical problem, both
the current and field would have to be expanded in asymptotic fields
as in section IX.

\section{conclusion}

The purpose of this work is to understand the relation between light-front
and more conventional formulations of quantum field theory.
Most of the conclusions are based on known properties of free fields
and properties of axiomatic formulations of field theories.

\bigskip

The first question is whether the vacuum of light front quantum field
theories is trivial.  

\begin{itemize}

\item[] Each theory has one vacuum.  It is a linear functional on the
  Heisenberg field algebra generated by covariant local fields smeared
  with Schwartz test functions in four variables.  The Wightman
  functions, which define the dynamics, are constructed by applying
  the vacuum functional to products of fields in the Heisenberg field
  algebra.  In quantum field theory the Wightman functions are the
  kernel of the Hilbert space inner product.  The covariance of the
  fields and invariance of the vacuum functional define a unitary
  representation of the Poincar\'e group on this representation of the
  Hilbert space.  Theories that have the same Wightman functions
  preserve the Hilbert space inner product and are unitarily
  equivalent.  Vacuum functionals that result in different
  Wightman functions do not preserve the Hilbert space inner product and
  define unitarily inequivalent theories.
  
\item[] Different vacuum functionals can agree on a sub-algebra of the
  Heisenberg algebra without agreeing on the whole algebra.  In
  section V and IX it was shown that vacuum functionals for free and
  asymptotic fields theories are defined and agree on the light-front
  Fock algebra.  In (\ref{ir22}) and (\ref{HR6}) it was shown that
  Wightman functions for
  the free and asymptotic Heisenberg field algebras could be expressed
  as vacuum expectation values of elements of sub-algebras of the
  light-front Fock algebra (The test functions in the light-front Fock
  algebra are restricted to Schwartz functions in the light-front
  coordinates that have vanishing $x^-$ derivative).  All of the
  vacuum functionals agree on this sub-algebra since they agree
  on the full light front Fock algebra.

  The assumed irreducibility of the asymptotic fields means that the
  Heisenberg fields can be expanded in a series of normal products of
  asymptotic fields; the vacuum does not change.  It then follows that
  the Wightman functions of the field theory can be formally expressed
  in terms of vacuum expectation values of the asymptotic fields which
  can in turn be expressed in terms vacuum expectation values of
  elements of a sub-algebra of the light-front Fock algebra.  Any of
  the vacuum functionals have the same value when applied to elements
  of this sub-algebra and can be used to evaluate the Wightman
  functions of the theory.

  The asymptotic fields generate a complete set of one-body and
  many-body scattering states that diagonailze the field theoretic
  Hamiltonian.  The masses of these fields encode the mass and spin
  spectrum of the field theoretic Poincar\'e generators.  Haag-Ruelle
  scattering theory defines the asymptotic fields in terms of
  Heisenberg fields that satisfy the Wightman axioms.  The dynamical
  difficulties appear in the inverse problem of how to define the
  Heisenberg algebra satisfying these axioms in terms of the algebra
  of asymptotic fields.  This is where all of the problems with
  renormalization appear.  The relevant
  expansion is given in \cite{Glaser:1957} where the coefficients of
  the expansion are only sensitive to the asymptotic fields on their
  mass shell. The input to the expansion is nested commutators
  of the Heisenberg fields with asymptotic fields. 
    
  This suggest that the free field Fock vacuum can be used provided the
  dynamical generators can be expressed as well-defined operators in the
  light-front Fock algebra.  In this case the light-front
  Hamiltonian determines 
  the dynamical sub-algebra of the light-front Fock algebra. 
   
\end{itemize}

\noindent The next question is the problem of inequivalent
representations of the canonical commutation relations.

\begin{itemize} 
  
\item[] As discussed above the Wightman functions for different
  theories can be expressed as vacuum expectation values of elements
  of a sub-algebra of the light-front Fock algebra.  While vacuum
  expectation values of elements of the sub-algebra are independent of
  the choice of vacuum functional, functionals applied to different
  sub-algebras are associcted with Wightman functions for different
  theories. The sub-algebras define the dynamics; they contain all of
  the information about the mass and spin spectrum of the interacting
  theory.  Inequivalent theories are associated with different
  sub-algebras of the light-front Fock algebra.  There is no relation
  between the vacuum expectation values of different sub-algebras of
  the light-front Fock algebra.  This is independent of the vacuum
  functional used to compute them.

\item[] The restriction of the fields to the light front has no
  dynamical information.  In addition fields on the light front are
  still operator valued distributions so they must first be smeared
  with suitable test functions.  Dynamical information enters when the
  test functions are restricted to the test functions constructed by
  mapping the four-dimensional test functions to the light front.
  This is demonstrated in (\ref{ir25}) and (\ref{ir26}) relating the
  two-point functions to two point functions on the light front.  The
  mapping encodes the correct approach to the light front in different
  theories.

\end{itemize}

\noindent The initial value problem

\begin{itemize}

\item[] The mapping from the local Heisenberg field algebra to the
  sub-algebra of the light-front Fock algebra maps covariant
  Poincar\'e generators to light-front Poincar\'e generators (see
  (\ref{ir24})).  For theories with a mass gap there is a dense set of
  vectors in the Hilbert space generated by the Heisenberg field
  algebra that gets mapped into a set of vectors generated by a
  sub-algebra of the light-front Fock algebra where the power series
  for the $x^+$ evolution operator converges (\ref{iv10}).  This is because the
  mapping to the sub-algebra enforces boundary conditions that
  eliminate light-like momenta.
  
\end{itemize}  

\noindent The problem of rotational covariance

\begin{itemize}

\item[] The Hamiltonian formulation of light-front quantum field
  theory defines infinitesimal generators of the Poincar\'e group by
  integrating the energy momentum and angular momentum tensors over
  the light front.  This results in formal expressions for the
  Poincar\'e generators expressed in term of fields on the light front
  and derivatives of fields tangent to the light front.  These formal
  expression are classical.  When the fields are replaced by operator
  valued distributions the expression for the dynamical generators
  become ill-defined.  These have to be renormalized in a manner that
  preserves the commutation relations and agrees with the
  representation obtained by mapping the covariant representations of
  the generators to the light-front Fock algebra
  (see (\ref{ir24})).
  
\item[] The light-front Hamiltonian, $P^-$ and the generators of the
  kinematic subgroup form a closed Lie algebra which does not uniquely
  determine the transverse rotation properties of the theory.  The
  light-front Hamiltonian is only defined after the 
  formal expression for the light-front Hamiltonian is renormalized.
  
\item[] The transverse rotational properties can be fixed by either
  replacing the light-front Hamiltonian by one transverse rotation
  generator or by demonstrating the invariance of the predictions of
  the theory by changing the orientation of the light front (see
  (\ref{RI12})).  In the first case there are non-linear constraints
  on acceptable transverse rotation generators while in the second
  case the equivalence requires matching correct scattering asymptotic
  conditions in theories with different light fronts.  This reduces to
  the problem of assigning the total spin to degenerate states with
  the same magnetic quantum numbers.

\item[] In light-front Hamiltonian forms of quantum field theories the
  problem of rotational invariance is connected with the problem of
  how to consistently renormalize both the $p\to \infty$ and $p^+ \to
  0$ divergences outside of perturbation theory.

\item[] In perturbation theory the rotational properties of the
  light-front formulation follow from the mapping from the Heisenberg
  algebra to the sub-algebra of the light-front Fock algebra.  In
  perturbation theory there are no composite states; the spins are
  fixed by the spinor representations of the fields.  The rotational
  properties can be determined by comparing light-front perturbation
  theory to covariant perturbation theory.  The problem with
  rotational covariance is related to the problem of assigning spins
  for degenerate composite states.  This is relevant outside of
  perturbation theory.

\item[] Rotational covariance remains a problem for non-perturbative
  applications.
    
\end{itemize}  

\noindent The problem of zero modes

\begin{itemize}

\item[] The light-front Hamiltonian and the kinematic generators do
  not fix the transverse rotational properties of the theory.  This
  means that in order to get a light-front dynamics
  that is equivalent to the covariant form of the theory,
  additional information that is not contained in the light-front
  Hamiltonian is needed.

\item[] Rotational invariance is equivalent to invariance with respect
  to changing the orientation of the light front.  The formal
  expressions for the Poincar\'e generators constructed using
  Noether's theorem are generally ill-defined.  They have both
  $p\to \infty$ and $p^+ \to 0$ divergences.  These divergence mix under
  changing orientation of the light front.  This means that the in
  order ensure rotational covariance the $p\to \infty$ and $p^+ \to 0$
  renormalizations cannot be done independently.  This means that some
  of the information missing from the light-front Hamiltonian is how
  to consistently treat the $p\to \infty$ and $p^+ \to 0$
  renormalization.  Similar problems occur with the space reflection
  operator, which already appears in $1+1$ dimensional applications.

\item[] This indicates that the renormalization of the $p^+\to 0$
  divergences and $p\to \infty$ divergences in the light-front 
  Hamiltonian are constrained by both rotational covariance and space
  reflection invariance.  This requires information that is not
  contained in the formal expressions for the light-front Hamiltonian.
  A complete treatment requires a non-perturbative renormalization.

\item[] In the perturbative case the connection with covariant
  formulations can be used to determine the needed $p^+\to 0$
  corrections to light-front perturbation theory \cite{Mannheim:2020rod}.
  
\end{itemize}  

\noindent The problem of spontaneous symmetry breaking 

\begin{itemize}
  
\item[] The signal for spontaneous symmetry breaking is the presence of a
  massless Goldstone boson in the mass spectrum.  While it is usually expressed
  in terms of the charge operator, the charge operator is the current
  density evaluated at a point $t=0$, $\mathbf{p}=0$ in a mixed time
  momentum representation.  It is an operator valued distribution and
  there is no reason to expect that it exists.  There is an unambiguous
  non-perturbative condition for the presence of a Goldstone boson that involves
  that vacuum expectation value of the commutator of a field and a
  current density integrated over a ball of finite volume in the limit that
  the volume $\to \infty$.  This works because locality eventually cuts off
  the integral when the ball gets sufficiently large, so there is
  never an integral over infinite volume.  This cannot be done with the
  corresponding light-front charge operator because locality is not
  available to cutoff the integral.   

\item[] Vacuum expectation values of the commutator of a cutoff
  current at fixed time with a field smeared over a compact region can
  be mapped to the light-front sub-algebra.  The non-vanishing of the
  light-front vacuum expectation values of operator in the limit of
  sufficiently large cutoff will be evidence of a Goldstone boson.
  While this can in principle be formulated in the light front
  sub-algebra, it is not the same as using the light-front charge
  operator.

\item[] Because spontaneous symmetry breaking is a
  non-pertrubative dynamical phenomena,  it is not surprising that it does not
  manifest itself on the light-front Fock algebra.  This consistent
  with treatments that formulate spontaneous symmetry breaking
  in terms of light-front limits of Heisenberg fields \cite{Lenz:2000} or
  in term of commutators with dynamical generators \cite{Beane:2013oia}.

\end{itemize}

The covariant and light-front formulations are two equivalent
representations of quantum field theory.  The vacuum in both theories
is the same and non-trivail, however when the fields are properly
mapped to the light front, the vacuum is unitarily equivalent to the
vacuum any free field vacuum.  This means that the vacuum can be
treated as a trivial vacuum for calculations on the light-front.
Noether's theorem on the light front (see Appendix 2) gives formal
expressions for the Poincar\'e generators as operators in the
light-front Fock algebra.  The formal expression for these operators
are ill-defined and require a consistent renormalization.  These
operators generate automorphisms in the light-front Fock algebra that
can be used to build the $x^+$ dependence of the fields.  

An important advantage of the light-front representation over the
covariant representation is that it has a natural Hamiltonian formulation,
which implies that many non-perturbative problems can be formally reduced
to linear algebra.  What is missing in the light-front case is a
non-perturbative way to renormalize the theory in a manner that is
consistent with rotational covariance and space reflection symmetry.
Assuming that this program can be performed,  this work suggests that
the vacuum can still be treated as trivial.  Issues with the dynamics
and 0 modes appear in the process of removing the divergences and
renormalizing the theory consistent with rotational covariance and
space reflection invariance.  

This research was
supported by the US Department of Energy, Office of Science, 
grant number DE-SC0016457.  The author would like to thank John Collins, Stan Brodsky, , 
Chueng Ji, and Peter Lowdon for useful comments on this work.

\section{Appendix 1} 

Light front conventions.  The light-front components of the spacetime coordinate
$x$ are

\[
x^{\pm} = x^0\pm x^3 = - (x_0\mp x_3) \qquad \mathbf{x}_{\perp}= (x_1,x_2)
\]
\[
x^0 = {1 \over 2}(x^++x^-) \qquad 
x^3 = {1 \over 2}(x^+-x^-).
\]
With these conventions the Lorentz invariant scalar product of
two four vectors is 
\[
x\cdot p = \mathbf{x}_{\perp}\cdot \mathbf{p}_{\perp}
- {1 \over 2} (x^-p^+ + x^+p^-).
\]

The covariant components of $x^{\mu}$ are
\[
x_+ = -{1 \over 2} x^- \qquad x_- = -{1 \over 2} x^+
\]
so
\[
x_+ x^+ + x_-x^- + x_1 x^1 + x_2 x^2 =
-{1 \over 2} x^+ x^- - {1 \over 2} x^- x^+ + \mathbf{x}_{\perp}\cdot \mathbf{x}_{\perp} = - x^{02} + \mathbf{x}\cdot \mathbf{x}.
\]

The 4-volume element and partial derivatives are
\[
d^4 x = {1 \over 2} dx^+ dx^- d^2\mathbf{x}_{\perp}
\]
\[
{\partial \over \partial x^0} = {\partial x^+\over \partial
  x^0}{\partial \over \partial x^+} + {\partial x^-\over \partial
  x^0}{\partial \over \partial x^-} = {\partial \over \partial
  x^+}+{\partial \over \partial x^-}
\qquad
{\partial \over \partial x^3} = {\partial x^+\over \partial
  x^3}{\partial \over \partial x^+} + {\partial x^-\over \partial
  x^3}{\partial \over \partial x^-} {\partial \over \partial
  x^+}-{\partial \over \partial x^-}.
\] 
With these conventions the Lagrangian density for a scalar field theory
has the form
\[
{\cal L} =
2 \partial_+(x)  \phi \partial_- \phi(x)
-{1 \over 2} \partial_i(x)  \phi \partial_i \phi(x)
-{m^2 \over 2} \phi^2 (x)
-V(\phi(x)).
\]

\section{Appendix 2} 

Poincar\'e invariance of the action leads to formal expressions for  
the infinitesimal generators of the Poincar\'e group using Noether's theorem.
The conserved currents are the energy momentum and angular momentum tensors
\beq
T^{\mu\nu} (x)= \eta^{\mu \nu}{\cal L} -
({\partial {\cal L} \over \partial (\partial_{\mu}\phi(x))})
\partial_{\alpha} \phi(x) \eta^{\alpha\nu} ) 
\label{A2:1}
\eeq
\beq
M^{\mu\alpha\beta}= T^{\mu\beta}x^\alpha - T^{\mu\alpha}x^\beta .
\label{A2:2}
\eeq
Poincar\'e invariance of the action and Lagrange's equation implies that these currents are conserved:
\beq
\partial_{\mu}T^{\mu\nu}(x) = 0
\qquad \mbox{and} \qquad
\partial_{\mu}M^{\mu\alpha \beta}(x) =0.
\label{A2:4}
\eeq
For a scalar field with Lagrangian density
\beq
{\cal L}(x) = -{1 \over 2} \partial^{\mu}\phi(x)\partial_{\mu}\phi(x)
-{m^2 \over 2}\phi(x)^2 - V(\phi (x)).
\label{A2:5}
\eeq
The field equations can be derived from the principle of stationary action
\beq
-\partial^{\mu}\partial_{\mu}\phi(x) + m^2 \phi(x) + {\partial V(\phi(x)) \over \partial \phi(x)} =0.
\label{A2:6}
\eeq
The Lagrangian density and field equations can be expressed in terms of light front variables as
\beq
{\cal L}(x) =  2\partial_{-}\phi(x)\partial_{+}\phi(x)
-{1 \over 2} \partial^{i}\phi(x)\partial_{i}\phi(x)
-{m^2 \over 2}\phi(x)^2 - V(\phi (x)).
\label{A2:11}
\eeq
and
\beq
4\partial_{-}\partial_{+}\phi(x)
-\partial^{i}\partial_{i}\phi(x)
+ m^2 \phi(x) + {\partial V(\phi(x)) \over \partial \phi(x)} =0
\label{A2:12}
\eeq

The Noether charges are constructed by integrating the $+$ component
of Noether currents over the $x^+=0$ light front where the $+$ components
of the currents are
\beq
T^{++} =
4 :\partial_- \phi(x) \partial_- \phi(x):
\label{A2:13}
\eeq
\beq
T^{+-} = 
-2 {\cal L} + 4 \partial_- \phi(x)) \partial_+ \phi(x) = 
:\pmb{\nabla}_{\perp} \phi(x) \cdot
\pmb{\nabla}_{\perp} \phi(x): + {m^2} :\phi(x)^2: + 2:V(\phi(x)):
\label{A2:14}
\eeq
\beq
T^{+i} =
-2 :\partial_- \phi(x) \partial_i \phi(x):
\label{A2:15}
\eeq
and
\beq
M^{+\alpha \beta}(x)= x^\alpha T^{+\beta}(x) -x^\beta T^{+\alpha}(x) =
x^\alpha T^{+\beta}(x) -x^\beta T^{+\alpha}(x)
\label{A2:16}
\eeq
Using these expressions the light front Poincar\'e generators can be
expressed in terms of the algebra of field operators restricted to the light front.  Note that all derivatives in these expression are normal to the
light front:
\beq
P^+ =
\int_{x^+=0} {d\mathbf{x}_{\perp} d x^- \over 2} T^{++}(x) =
\int_{x^+=0} {d\mathbf{x}_{\perp} d x^- \over 2}4 :\partial_- \phi(x) \partial_- \phi(x): 
\label{A2:19}
\eeq
\beq
P^i =\int_{x^+=0} {d\mathbf{x}_{\perp} d x^- \over 2}
T^{+i}  =
-\int_{x^+=0} {d\mathbf{x}_{\perp} d x^- \over 2}
2 :\partial_- \phi(x) \partial_i \phi(x):  \qquad i \in \{1,2\}
\label{A2:20}
\eeq
\beq
{E}^i = \int_{x^+=0} {d\mathbf{x}_{\perp} d x^- \over 2} T^{++}x^i =
\int_{x^+=0} {d\mathbf{x}_{\perp} d x^- \over 2} 4 :\partial_- \phi(x) \partial_- \phi(x):x^i
\label{A2:21}
\eeq
\[
J^3 = \int_{x^+=0} {d\mathbf{x}_{\perp} d x^- \over 2}
\left(
x^1 T^{+2}(x) - x^2 T^{+1} \right ) =
\]
\beq
\int_{x^+=0} {d\mathbf{x}_{\perp} d x^- \over 2}
\left(
- 2 x^1 :\partial_- \phi(x) \partial_2 \phi(x): + 
2 x^2 :\partial_- \phi(x) \partial_1 \phi(x):\right ) 
\label{A2:22}
\eeq
\beq
K^3 = \int_{x^+=0} {d\mathbf{x}_{\perp} d x^- \over 2} 
T^{++}(x) x^- =
\int_{x^+=0} {d\mathbf{x}_{\perp} d x^- \over 2} 
4 :\partial_- \phi(x) \partial_- \phi(x): x^-
\label{A2:23}
\eeq
The dynamical generators are:
\beq
P^- = \int_{x^+=0} {d\mathbf{x}_{\perp} d x^- \over 2} T^{+-}(x) =
\label{A2:17}
\eeq
\beq
\int_{x^+=0} {d\mathbf{x}_{\perp} d x^- \over 2}
\left (
 :\pmb{\nabla}_{\perp} \phi(x) \cdot
\pmb{\nabla}_{\perp} \phi(x): + {m^2} :\phi(x)^2: + 2 :V(\phi(x)):
\right )
\label{A2:18}
\eeq
\[
J^1 =
\int_{x^+=0} {d\mathbf{x}_{\perp} d x^- \over 4}
\left(
x^2 (T^{++}(x)- T^{+-}) + x^- T^{+2} \right ) =
\]
\beq
\int_{x^+=0} {d\mathbf{x}_{\perp} d x^- \over 4}
\left(
x^2 (4 :\partial_- \phi(x) \partial_- \phi(x): 
-:\pmb{\nabla}_{\perp} \phi(x) \cdot
\pmb{\nabla}_{\perp} \phi(x): - {m^2} :\phi(x)^2: - 2:V(\phi(x)):
) - x^- 2 :\partial_- \phi(x) \partial_2 \phi(x): \right )
\label{A2:24}
\eeq
\[
J^2 = -\int_{x^+=0} {d\mathbf{x}_{\perp} d x^- \over 4} 
\left(
x^- T^{+1}(x) + x^1 (T^{++}- T^{+-})) \right ) =
\]
\beq
\int_{x^+=0} {d\mathbf{x}_{\perp} d x^- \over 4} 
\left(
- x^- 2 :\partial_- \phi(x) \partial_1 \phi(x):
\right .
\left .
- x^1 (4 :\partial_- \phi(x) \partial_- \phi(x): -  (
 :\pmb{\nabla}_{\perp} \phi(x) \cdot
\pmb{\nabla}_{\perp} \phi(x): + {m^2} :\phi(x)^2: + 2:V(\phi(x)):
) \right )
\label{A2:25}
\eeq
For the case of free fields these operators can be expressed in terms of
light-front creation and annihilation operators
\beq
P^+=
\int d\tilde{\mathbf{p}} \theta (p^+)
a^{\dagger}(\tilde{\mathbf{p}})p^+ a(\tilde{\mathbf{p}})
\label{A2:26}
\eeq
\beq
P^- =
\int d\tilde{\mathbf{p}} \theta (p^+)
a^{\dagger}(\tilde{\mathbf{p}}){\mathbf{p}_{\perp}+m^2 \over p^+}
a(\tilde{\mathbf{p}})
\label{A2:27}
\eeq
\beq
P^i=
\int d\tilde{\mathbf{p}} \theta (p^+)
a^{\dagger}(\tilde{\mathbf{p}})p^i a(\tilde{\mathbf{p}})
\label{A2:28}
\eeq

\beq
E^i = \int d\tilde{\mathbf{p}} \theta (p^+)
a^{\dagger}(\tilde{\mathbf{p}})(-i p^+{\partial \over \partial p^i}) a(\tilde{\mathbf{p}})
\label{A2:28a}
\eeq
\beq
K^3 \int d\tilde{\mathbf{p}} \theta (p^+)
a^{\dagger}(\tilde{\mathbf{p}})
(i \{p^+,{\partial \over \partial p^+}\} )
a(\tilde{\mathbf{p}})
\label{A2:28b}
\eeq
\beq
J^3 =
\int d\tilde{\mathbf{p}} \theta (p^+)
a^{\dagger}(\tilde{\mathbf{p}})
(x^2(i\partial_{p_1}) - x^1(i\partial_{p_2}))  
a(\tilde{\mathbf{p}})
\label{A2:29}
\eeq
\[
{1 \over 2}\int d\tilde{\mathbf{p}} \theta (p^+)
a^{\dagger}(\tilde{\mathbf{p}}) =
(
\{ {\mathbf{p}^2_{\perp}+m^2 \over p^+}, (i \partial_{pi})\} )
a(\tilde{\mathbf{p}})
\]
\beq
J^1 =
\int d\tilde{\mathbf{p}}
\theta (p^+)
a^{\dagger}(\tilde{\mathbf{p}})
\left (
{1 \over 2} p^+(i \partial_{p_2})
- {1 \over 4}\{ {\mathbf{p}_{\perp}^2 + m^2 \over p^+},i \partial_{p_2} \}
- {1 \over 2} (2i p^2 \partial_{p+}) 
\right )
a(\tilde{\mathbf{p}})
\label{A2:30}
\eeq
\beq
J^2 =
\int d\tilde{\mathbf{p}}
\theta (p^+)
a^{\dagger}(\tilde{\mathbf{p}})
\left (
-{1 \over 2} p^+(i \partial_{p_1})
+ {1 \over 4}\{ {\mathbf{p}_{\perp}^2 + m^2 \over p^+},i \partial_{p_1} \}
+ {1 \over 2} (2i p^1 \partial_{p+}) 
\right )
a(\tilde{\mathbf{p}}).
\label{A2:30}
\eeq
The three dynamical generators have a dependence on the mass $m$ and only involve derivatives tangent to the light front.

\bibliography{collins.bib}

\begin{thebibliography}{84}
\expandafter\ifx\csname natexlab\endcsname\relax\def\natexlab#1{#1}\fi
\expandafter\ifx\csname bibnamefont\endcsname\relax
  \def\bibnamefont#1{#1}\fi
\expandafter\ifx\csname bibfnamefont\endcsname\relax
  \def\bibfnamefont#1{#1}\fi
\expandafter\ifx\csname citenamefont\endcsname\relax
  \def\citenamefont#1{#1}\fi
\expandafter\ifx\csname url\endcsname\relax
  \def\url#1{\texttt{#1}}\fi
\expandafter\ifx\csname urlprefix\endcsname\relax\def\urlprefix{URL }\fi
\providecommand{\bibinfo}[2]{#2}
\providecommand{\eprint}[2][]{\url{#2}}

\bibitem[{\citenamefont{Chang and Ma}(1969)}]{Chang1069}
\bibinfo{author}{\bibfnamefont{S.-J.} \bibnamefont{Chang}} \bibnamefont{and}
  \bibinfo{author}{\bibfnamefont{S.-K.} \bibnamefont{Ma}},
  \bibinfo{journal}{Phys. Rev.} \textbf{\bibinfo{volume}{180}},
  \bibinfo{pages}{1506} (\bibinfo{year}{1969}),
  \urlprefix\url{https://link.aps.org/doi/10.1103/PhysRev.180.1506}.

\bibitem[{\citenamefont{Kogut and Soper}(1970)}]{Soper}
\bibinfo{author}{\bibfnamefont{J.~B.} \bibnamefont{Kogut}} \bibnamefont{and}
  \bibinfo{author}{\bibfnamefont{D.~E.} \bibnamefont{Soper}},
  \bibinfo{journal}{Phys. Rev. D} \textbf{\bibinfo{volume}{1}},
  \bibinfo{pages}{2901} (\bibinfo{year}{1970}),
  \urlprefix\url{https://link.aps.org/doi/10.1103/PhysRevD.1.2901}.

\bibitem[{\citenamefont{Chang et~al.}(1973)\citenamefont{Chang, Root, and
  Yan}}]{Chang:1972xt}
\bibinfo{author}{\bibfnamefont{S.-J.} \bibnamefont{Chang}},
  \bibinfo{author}{\bibfnamefont{R.~G.} \bibnamefont{Root}}, \bibnamefont{and}
  \bibinfo{author}{\bibfnamefont{T.-M.} \bibnamefont{Yan}},
  \bibinfo{journal}{Phys.Rev.} \textbf{\bibinfo{volume}{D7}},
  \bibinfo{pages}{1133} (\bibinfo{year}{1973}).

\bibitem[{\citenamefont{Chang and Yan}(1973)}]{Yan:1973.2}
\bibinfo{author}{\bibfnamefont{S.-J.} \bibnamefont{Chang}} \bibnamefont{and}
  \bibinfo{author}{\bibfnamefont{T.-M.} \bibnamefont{Yan}},
  \bibinfo{journal}{Phys. Rev. D} \textbf{\bibinfo{volume}{7}},
  \bibinfo{pages}{1147} (\bibinfo{year}{1973}),
  \urlprefix\url{https://link.aps.org/doi/10.1103/PhysRevD.7.1147}.

\bibitem[{\citenamefont{Yan}(1973{\natexlab{a}})}]{Yan:1973.3}
\bibinfo{author}{\bibfnamefont{T.-M.} \bibnamefont{Yan}},
  \bibinfo{journal}{Phys. Rev. D} \textbf{\bibinfo{volume}{7}},
  \bibinfo{pages}{1760} (\bibinfo{year}{1973}{\natexlab{a}}),
  \urlprefix\url{https://link.aps.org/doi/10.1103/PhysRevD.7.1760}.

\bibitem[{\citenamefont{Yan}(1973{\natexlab{b}})}]{Yan:1973.4}
\bibinfo{author}{\bibfnamefont{T.-M.} \bibnamefont{Yan}},
  \bibinfo{journal}{Phys. Rev. D} \textbf{\bibinfo{volume}{7}},
  \bibinfo{pages}{1780} (\bibinfo{year}{1973}{\natexlab{b}}),
  \urlprefix\url{https://link.aps.org/doi/10.1103/PhysRevD.7.1780}.

\bibitem[{\citenamefont{Vary et~al.}(2010)\citenamefont{Vary, Honkanen, Li,
  Maris, Brodsky, Harindranath, {de Teramond}, Sternberg, Ng, and
  Yang}}]{VARY201064}
\bibinfo{author}{\bibfnamefont{J.}~\bibnamefont{Vary}},
  \bibinfo{author}{\bibfnamefont{H.}~\bibnamefont{Honkanen}},
  \bibinfo{author}{\bibfnamefont{J.}~\bibnamefont{Li}},
  \bibinfo{author}{\bibfnamefont{P.}~\bibnamefont{Maris}},
  \bibinfo{author}{\bibfnamefont{S.}~\bibnamefont{Brodsky}},
  \bibinfo{author}{\bibfnamefont{A.}~\bibnamefont{Harindranath}},
  \bibinfo{author}{\bibfnamefont{G.}~\bibnamefont{{de Teramond}}},
  \bibinfo{author}{\bibfnamefont{P.}~\bibnamefont{Sternberg}},
  \bibinfo{author}{\bibfnamefont{E.}~\bibnamefont{Ng}}, \bibnamefont{and}
  \bibinfo{author}{\bibfnamefont{C.}~\bibnamefont{Yang}},
  \bibinfo{journal}{Nuclear Physics B - Proceedings Supplements}
  \textbf{\bibinfo{volume}{199}}, \bibinfo{pages}{64} (\bibinfo{year}{2010}),
  ISSN \bibinfo{issn}{0920-5632}, \bibinfo{note}{proceedings of the
  International Workshop Light Cone 2009 (LC2009): Relativistic Hadronic and
  Particle Physics},
  \urlprefix\url{https://www.sciencedirect.com/science/article/pii/S0920563210000289}.

\bibitem[{\citenamefont{Stone}(1930)}]{MHStone1}
\bibinfo{author}{\bibfnamefont{M.~H.} \bibnamefont{Stone}},
  \bibinfo{journal}{Proceedings of the National Academy of Sciences of the
  United States of America} \textbf{\bibinfo{volume}{vol. 16}},
  \bibinfo{pages}{172–175} (\bibinfo{year}{1930}).

\bibitem[{\citenamefont{von Neumann}(1931)}]{JvNeumann2}
\bibinfo{author}{\bibfnamefont{J.}~\bibnamefont{von Neumann}},
  \bibinfo{journal}{Mathematische Annalen} \textbf{\bibinfo{volume}{vol. 104}},
  \bibinfo{pages}{570–578} (\bibinfo{year}{1931}).

\bibitem[{\citenamefont{von Neumann}(1932)}]{JvNeumann3}
\bibinfo{author}{\bibfnamefont{J.}~\bibnamefont{von Neumann}},
  \bibinfo{journal}{Annals of Mathematics} \textbf{\bibinfo{volume}{33}},
  \bibinfo{pages}{567–573} (\bibinfo{year}{1932}).

\bibitem[{\citenamefont{Haag}(1955)}]{Haag:1955ev}
\bibinfo{author}{\bibfnamefont{R.}~\bibnamefont{Haag}},
  \bibinfo{journal}{Mat-Fys. Medd. K. Danske Vidensk. Selsk.}
  \textbf{\bibinfo{volume}{29}}, \bibinfo{pages}{1} (\bibinfo{year}{1955}).

\bibitem[{\citenamefont{Schlieder and Seiler}(1972)}]{Schlieder:1972qr}
\bibinfo{author}{\bibfnamefont{S.}~\bibnamefont{Schlieder}} \bibnamefont{and}
  \bibinfo{author}{\bibfnamefont{E.}~\bibnamefont{Seiler}},
  \bibinfo{journal}{Commun. Math. Phys.} \textbf{\bibinfo{volume}{25}},
  \bibinfo{pages}{62} (\bibinfo{year}{1972}).

\bibitem[{\citenamefont{Weinberg}(2009)}]{Weinberg:2009ca}
\bibinfo{author}{\bibfnamefont{S.}~\bibnamefont{Weinberg}}
  (\bibinfo{year}{2009}), \eprint{0903.0568}.

\bibitem[{\citenamefont{Weinberg}(1967)}]{PhysRev.158.1638}
\bibinfo{author}{\bibfnamefont{S.}~\bibnamefont{Weinberg}},
  \bibinfo{journal}{Phys. Rev.} \textbf{\bibinfo{volume}{158}},
  \bibinfo{pages}{1638} (\bibinfo{year}{1967}),
  \urlprefix\url{https://link.aps.org/doi/10.1103/PhysRev.158.1638}.

\bibitem[{\citenamefont{Araki}(1964)}]{Araki:1964}
\bibinfo{author}{\bibfnamefont{H.}~\bibnamefont{Araki}}, \bibinfo{journal}{J.
  Math. Phys.} \textbf{\bibinfo{volume}{1}}, \bibinfo{pages}{492}
  (\bibinfo{year}{1964}).

\bibitem[{\citenamefont{Coester and Haag}(1960)}]{PhysRev.117.1137}
\bibinfo{author}{\bibfnamefont{F.}~\bibnamefont{Coester}} \bibnamefont{and}
  \bibinfo{author}{\bibfnamefont{R.}~\bibnamefont{Haag}},
  \bibinfo{journal}{Phys. Rev.} \textbf{\bibinfo{volume}{117}},
  \bibinfo{pages}{1137} (\bibinfo{year}{1960}).

\bibitem[{\citenamefont{Karmanov}(1976{\natexlab{a}})}]{Karmanov1}
\bibinfo{author}{\bibfnamefont{V.}~\bibnamefont{Karmanov}},
  \bibinfo{journal}{Sov. Phys. JETP} \textbf{\bibinfo{volume}{44}},
  \bibinfo{pages}{210} (\bibinfo{year}{1976}{\natexlab{a}}).

\bibitem[{\citenamefont{Karmanov}(1976{\natexlab{b}})}]{Karmanov2}
\bibinfo{author}{\bibfnamefont{V.}~\bibnamefont{Karmanov}},
  \bibinfo{journal}{Sov. Phys. JETP} \textbf{\bibinfo{volume}{48}},
  \bibinfo{pages}{598} (\bibinfo{year}{1976}{\natexlab{b}}).

\bibitem[{\citenamefont{Karmanov}(1979{\natexlab{a}})}]{Karmanov3}
\bibinfo{author}{\bibfnamefont{V.}~\bibnamefont{Karmanov}},
  \bibinfo{journal}{Sov. Phys. JETP} \textbf{\bibinfo{volume}{76}},
  \bibinfo{pages}{1884} (\bibinfo{year}{1979}{\natexlab{a}}).

\bibitem[{\citenamefont{Karmanov}(1979{\natexlab{b}})}]{Karmanov4}
\bibinfo{author}{\bibfnamefont{V.}~\bibnamefont{Karmanov}},
  \bibinfo{journal}{Sov. Phys. JETP} \textbf{\bibinfo{volume}{49}},
  \bibinfo{pages}{954} (\bibinfo{year}{1979}{\natexlab{b}}).

\bibitem[{\citenamefont{Karmanov}(1980)}]{Karmanov5}
\bibinfo{author}{\bibfnamefont{V.}~\bibnamefont{Karmanov}},
  \bibinfo{journal}{Nucl. Phys. B} \textbf{\bibinfo{volume}{166}},
  \bibinfo{pages}{378} (\bibinfo{year}{1980}).

\bibitem[{\citenamefont{Fuda}(1994)}]{Fuda:1994uv}
\bibinfo{author}{\bibfnamefont{M.}~\bibnamefont{Fuda}},
  \bibinfo{journal}{Annals Phys.} \textbf{\bibinfo{volume}{231}},
  \bibinfo{pages}{1} (\bibinfo{year}{1994}).

\bibitem[{\citenamefont{Fuda}(1990)}]{Fuda:1990}
\bibinfo{author}{\bibfnamefont{M.}~\bibnamefont{Fuda}},
  \bibinfo{journal}{Annals Phys.} \textbf{\bibinfo{volume}{197}},
  \bibinfo{pages}{265} (\bibinfo{year}{1990}).

\bibitem[{\citenamefont{Polyzou}(1999)}]{Polyzou:1999}
\bibinfo{author}{\bibfnamefont{W.~N.} \bibnamefont{Polyzou}},
  \bibinfo{journal}{Few Body Syst.} \textbf{\bibinfo{volume}{27}},
  \bibinfo{pages}{57} (\bibinfo{year}{1999}).

\bibitem[{\citenamefont{Maskawa and Yamawaki}(1976)}]{Maskawa:1976}
\bibinfo{author}{\bibfnamefont{T.}~\bibnamefont{Maskawa}} \bibnamefont{and}
  \bibinfo{author}{\bibfnamefont{K.}~\bibnamefont{Yamawaki}},
  \bibinfo{journal}{Prog. Theor. Phys.} \textbf{\bibinfo{volume}{56}},
  \bibinfo{pages}{270} (\bibinfo{year}{1976}).

\bibitem[{\citenamefont{Yamawaki}(1998)}]{Yamawaki:1998}
\bibinfo{author}{\bibfnamefont{K.}~\bibnamefont{Yamawaki}}
  (\bibinfo{year}{1998}), \eprint{hep-th/9802037}.

\bibitem[{\citenamefont{Choi and Ji}(1998)}]{choi:1998}
\bibinfo{author}{\bibfnamefont{H.-M.} \bibnamefont{Choi}} \bibnamefont{and}
  \bibinfo{author}{\bibfnamefont{C.-R.} \bibnamefont{Ji}},
  \bibinfo{journal}{Phys. Rev. D} \textbf{\bibinfo{volume}{58}},
  \bibinfo{pages}{071901(R)} (\bibinfo{year}{1998}),
  \urlprefix\url{https://link.aps.org/doi/10.1103/PhysRevD.58.071901}.

\bibitem[{\citenamefont{Leutwyler et~al.}(1970)\citenamefont{Leutwyler,
  Klauder, and Streit}}]{Leutwyler:1970wn}
\bibinfo{author}{\bibfnamefont{H.}~\bibnamefont{Leutwyler}},
  \bibinfo{author}{\bibfnamefont{J.~R.} \bibnamefont{Klauder}},
  \bibnamefont{and} \bibinfo{author}{\bibfnamefont{L.}~\bibnamefont{Streit}},
  \bibinfo{journal}{Nuovo Cim.} \textbf{\bibinfo{volume}{A66}},
  \bibinfo{pages}{536} (\bibinfo{year}{1970}).

\bibitem[{\citenamefont{Rohrlich}(1971)}]{Rohrlich:1971zz}
\bibinfo{author}{\bibfnamefont{F.}~\bibnamefont{Rohrlich}},
  \bibinfo{journal}{Acta Phys.Austriaca Suppl.} \textbf{\bibinfo{volume}{8}},
  \bibinfo{pages}{277} (\bibinfo{year}{1971}).

\bibitem[{\citenamefont{Nakanishi and Yabuki}(1977)}]{Nakanishi:1976yx}
\bibinfo{author}{\bibfnamefont{N.}~\bibnamefont{Nakanishi}} \bibnamefont{and}
  \bibinfo{author}{\bibfnamefont{H.}~\bibnamefont{Yabuki}},
  \bibinfo{journal}{Lett. Math. Phys.} \textbf{\bibinfo{volume}{1}},
  \bibinfo{pages}{371} (\bibinfo{year}{1977}).

\bibitem[{\citenamefont{Nakanishi and Yamawaki}(1977)}]{Nakanishi:1977}
\bibinfo{author}{\bibfnamefont{N.}~\bibnamefont{Nakanishi}} \bibnamefont{and}
  \bibinfo{author}{\bibfnamefont{K.}~\bibnamefont{Yamawaki}},
  \bibinfo{journal}{Nucl. Phys.} \textbf{\bibinfo{volume}{B122}},
  \bibinfo{pages}{15} (\bibinfo{year}{1977}).

\bibitem[{\citenamefont{Leutwyler and Stern}(1978)}]{Leutwyler:1977vy}
\bibinfo{author}{\bibfnamefont{H.}~\bibnamefont{Leutwyler}} \bibnamefont{and}
  \bibinfo{author}{\bibfnamefont{J.}~\bibnamefont{Stern}},
  \bibinfo{journal}{Ann. Phys.} \textbf{\bibinfo{volume}{112}},
  \bibinfo{pages}{94} (\bibinfo{year}{1978}).

\bibitem[{\citenamefont{Coester}(1992)}]{Coester:1992}
\bibinfo{author}{\bibfnamefont{F.}~\bibnamefont{Coester}},
  \bibinfo{journal}{Progress in Particle and Nuclear Physics}
  \textbf{\bibinfo{volume}{29}}, \bibinfo{pages}{1} (\bibinfo{year}{1992}).

\bibitem[{\citenamefont{Coester and Polyzou}(1994)}]{Coester:1993fg}
\bibinfo{author}{\bibfnamefont{F.}~\bibnamefont{Coester}} \bibnamefont{and}
  \bibinfo{author}{\bibfnamefont{W.}~\bibnamefont{Polyzou}},
  \bibinfo{journal}{Found. Phys.} \textbf{\bibinfo{volume}{24}},
  \bibinfo{pages}{387} (\bibinfo{year}{1994}), \eprint{nucl-th/9305005}.

\bibitem[{\citenamefont{Wilson et~al.}(1994)\citenamefont{Wilson, Walhout,
  Harindrarath, Zhang, Perry, and Glazek}}]{wilson:1994}
\bibinfo{author}{\bibfnamefont{K.~G.} \bibnamefont{Wilson}},
  \bibinfo{author}{\bibfnamefont{T.~S.} \bibnamefont{Walhout}},
  \bibinfo{author}{\bibfnamefont{A.}~\bibnamefont{Harindrarath}},
  \bibinfo{author}{\bibfnamefont{W.-M.} \bibnamefont{Zhang}},
  \bibinfo{author}{\bibfnamefont{R.~J.} \bibnamefont{Perry}}, \bibnamefont{and}
  \bibinfo{author}{\bibfnamefont{S.~D.} \bibnamefont{Glazek}},
  \bibinfo{journal}{Physical Review D} \textbf{\bibinfo{volume}{49}},
  \bibinfo{pages}{6720} (\bibinfo{year}{1994}).

\bibitem[{\citenamefont{Bylev}(1996)}]{Bylev:1996}
\bibinfo{author}{\bibfnamefont{A.}~\bibnamefont{Bylev}}, \bibinfo{journal}{J.
  Phys. G} \textbf{\bibinfo{volume}{22}}, \bibinfo{pages}{1553}
  (\bibinfo{year}{1996}).

\bibitem[{\citenamefont{Brodsky et~al.}(1998)\citenamefont{Brodsky, Pauli, and
  Pinsky}}]{Brodsky:1998}
\bibinfo{author}{\bibfnamefont{S.}~\bibnamefont{Brodsky}},
  \bibinfo{author}{\bibfnamefont{H.-C.} \bibnamefont{Pauli}}, \bibnamefont{and}
  \bibinfo{author}{\bibfnamefont{S.}~\bibnamefont{Pinsky}},
  \bibinfo{journal}{Physics Reports} \textbf{\bibinfo{volume}{301}},
  \bibinfo{pages}{299} (\bibinfo{year}{1998}).

\bibitem[{\citenamefont{Tsujimaru and Yamawaki}(1998)}]{Tsujimaru:1997jt}
\bibinfo{author}{\bibfnamefont{S.}~\bibnamefont{Tsujimaru}} \bibnamefont{and}
  \bibinfo{author}{\bibfnamefont{K.}~\bibnamefont{Yamawaki}},
  \bibinfo{journal}{Phys.Rev.} \textbf{\bibinfo{volume}{D57}},
  \bibinfo{pages}{4942} (\bibinfo{year}{1998}), \eprint{hep-th/9704171}.

\bibitem[{\citenamefont{Lenz}(2000)}]{Lenz:2000}
\bibinfo{author}{\bibfnamefont{F.}~\bibnamefont{Lenz}}, \bibinfo{journal}{Nucl.
  Phys. (Proc. Suppl.)} \textbf{\bibinfo{volume}{B90}}, \bibinfo{pages}{46}
  (\bibinfo{year}{2000}).

\bibitem[{\citenamefont{Heinzl}(2001)}]{Heinzl:2001}
\bibinfo{author}{\bibfnamefont{T.}~\bibnamefont{Heinzl}},
  \bibinfo{journal}{Lec. Notes Ohys.} \textbf{\bibinfo{volume}{572}},
  \bibinfo{pages}{55} (\bibinfo{year}{2001}), \eprint{hep-th/0008096}.

\bibitem[{\citenamefont{Burkardt et~al.}(2002)\citenamefont{Burkardt, Lenz, and
  Thies}}]{Burkhardt:2002}
\bibinfo{author}{\bibfnamefont{M.}~\bibnamefont{Burkardt}},
  \bibinfo{author}{\bibfnamefont{F.}~\bibnamefont{Lenz}}, \bibnamefont{and}
  \bibinfo{author}{\bibfnamefont{M.}~\bibnamefont{Thies}},
  \bibinfo{journal}{Phys. Rev. D} \textbf{\bibinfo{volume}{65}},
  \bibinfo{pages}{125002} (\bibinfo{year}{2002}),
  \urlprefix\url{https://link.aps.org/doi/10.1103/PhysRevD.65.125002}.

\bibitem[{\citenamefont{Srivastava and Brodsky}(2001)}]{brodsky_srivastava_1}
\bibinfo{author}{\bibfnamefont{P.~P.} \bibnamefont{Srivastava}}
  \bibnamefont{and} \bibinfo{author}{\bibfnamefont{S.~J.}
  \bibnamefont{Brodsky}}, \bibinfo{journal}{Phys. Rev. D}
  \textbf{\bibinfo{volume}{64}}, \bibinfo{pages}{045006}
  (\bibinfo{year}{2001}),
  \urlprefix\url{https://link.aps.org/doi/10.1103/PhysRevD.64.045006}.

\bibitem[{\citenamefont{Srivastava and Brodsky}(2002)}]{brodsky_2002}
\bibinfo{author}{\bibfnamefont{P.~P.} \bibnamefont{Srivastava}}
  \bibnamefont{and} \bibinfo{author}{\bibfnamefont{S.~J.}
  \bibnamefont{Brodsky}}, \bibinfo{journal}{Phys. Rev. D}
  \textbf{\bibinfo{volume}{66}}, \bibinfo{pages}{045019}
  (\bibinfo{year}{2002}),
  \urlprefix\url{https://link.aps.org/doi/10.1103/PhysRevD.66.045019}.

\bibitem[{\citenamefont{Heinzl}(2003)}]{Heinzl:2003}
\bibinfo{author}{\bibfnamefont{T.}~\bibnamefont{Heinzl}}
  (\bibinfo{year}{2003}), \eprint{hep-th/0310165}.

\bibitem[{\citenamefont{Ullrich and Werner}(2006)}]{Werner:2006}
\bibinfo{author}{\bibfnamefont{P.}~\bibnamefont{Ullrich}} \bibnamefont{and}
  \bibinfo{author}{\bibfnamefont{E.}~\bibnamefont{Werner}},
  \bibinfo{journal}{J. Phys. A} \textbf{\bibinfo{volume}{39}},
  \bibinfo{pages}{6057} (\bibinfo{year}{2006}).

\bibitem[{\citenamefont{Martinovi{\v{c}} and Grange}(2008)}]{lubo2008higgs}
\bibinfo{author}{\bibfnamefont{L.}~\bibnamefont{Martinovi{\v{c}}}}
  \bibnamefont{and} \bibinfo{author}{\bibfnamefont{P.}~\bibnamefont{Grange}},
  \bibinfo{journal}{Modern Physics Letters A} \textbf{\bibinfo{volume}{23}},
  \bibinfo{pages}{417} (\bibinfo{year}{2008}).

\bibitem[{\citenamefont{Bakker}(2011)}]{Bakker:2011zza}
\bibinfo{author}{\bibfnamefont{B.~L.} \bibnamefont{Bakker}},
  \bibinfo{journal}{Few Body Syst.} \textbf{\bibinfo{volume}{49}},
  \bibinfo{pages}{177} (\bibinfo{year}{2011}).

\bibitem[{\citenamefont{Choi and Ji}(2011)}]{Choi:2011xm}
\bibinfo{author}{\bibfnamefont{H.-M.} \bibnamefont{Choi}} \bibnamefont{and}
  \bibinfo{author}{\bibfnamefont{C.-R.} \bibnamefont{Ji}},
  \bibinfo{journal}{Phys. Lett. B} \textbf{\bibinfo{volume}{696}},
  \bibinfo{pages}{518} (\bibinfo{year}{2011}), \eprint{1101.3035}.

\bibitem[{\citenamefont{Choi and Ji}(2014)}]{Choi:2013ira}
\bibinfo{author}{\bibfnamefont{H.-M.} \bibnamefont{Choi}} \bibnamefont{and}
  \bibinfo{author}{\bibfnamefont{C.-R.} \bibnamefont{Ji}},
  \bibinfo{journal}{Few Body Syst.} \textbf{\bibinfo{volume}{55}},
  \bibinfo{pages}{435} (\bibinfo{year}{2014}), \eprint{1311.0552}.

\bibitem[{\citenamefont{Beane}(2013)}]{Beane:2013oia}
\bibinfo{author}{\bibfnamefont{S.~R.} \bibnamefont{Beane}},
  \bibinfo{journal}{Annals Phys.} \textbf{\bibinfo{volume}{337}},
  \bibinfo{pages}{111} (\bibinfo{year}{2013}), \eprint{1302.1600}.

\bibitem[{\citenamefont{Chabysheva and Hiller}(2014)}]{sofia:2014}
\bibinfo{author}{\bibfnamefont{S.~S.} \bibnamefont{Chabysheva}}
  \bibnamefont{and} \bibinfo{author}{\bibfnamefont{J.~R.}
  \bibnamefont{Hiller}}, \bibinfo{journal}{Annals of Physics}
  \textbf{\bibinfo{volume}{340}}, \bibinfo{pages}{188 } (\bibinfo{year}{2014}),
  ISSN \bibinfo{issn}{0003-4916}.

\bibitem[{\citenamefont{Herrmann and Polyzou}(2015)}]{Herrmann:2015dqa}
\bibinfo{author}{\bibfnamefont{M.}~\bibnamefont{Herrmann}} \bibnamefont{and}
  \bibinfo{author}{\bibfnamefont{W.~N.} \bibnamefont{Polyzou}},
  \bibinfo{journal}{Phys. Rev. D} \textbf{\bibinfo{volume}{91}},
  \bibinfo{pages}{085043} (\bibinfo{year}{2015}), \eprint{1502.01230}.

\bibitem[{\citenamefont{Brodsky et~al.}(2015)\citenamefont{Brodsky, {de
  Téramond}, Dosch, and Erlich}}]{BRODSKY20151}
\bibinfo{author}{\bibfnamefont{S.~J.} \bibnamefont{Brodsky}},
  \bibinfo{author}{\bibfnamefont{G.~F.} \bibnamefont{{de Téramond}}},
  \bibinfo{author}{\bibfnamefont{H.~G.} \bibnamefont{Dosch}}, \bibnamefont{and}
  \bibinfo{author}{\bibfnamefont{J.}~\bibnamefont{Erlich}},
  \bibinfo{journal}{Physics Reports} \textbf{\bibinfo{volume}{584}},
  \bibinfo{pages}{1} (\bibinfo{year}{2015}), ISSN \bibinfo{issn}{0370-1573},
  \bibinfo{note}{light-front holographic QCD and emerging confinement},
  \urlprefix\url{https://www.sciencedirect.com/science/article/pii/S0370157315002306}.

\bibitem[{\citenamefont{Hiller}(2016)}]{Hiller:2016}
\bibinfo{author}{\bibfnamefont{J.}~\bibnamefont{Hiller}},
  \bibinfo{journal}{Prog. Part. Nucl. Phys.} \textbf{\bibinfo{volume}{90}},
  \bibinfo{pages}{75} (\bibinfo{year}{2016}), \eprint{hep-th/1606.08348}.

\bibitem[{\citenamefont{Ji}(2017)}]{Ji:2017vfu}
\bibinfo{author}{\bibfnamefont{C.-R.} \bibnamefont{Ji}}, \bibinfo{journal}{Few
  Body Syst.} \textbf{\bibinfo{volume}{58}}, \bibinfo{pages}{42}
  (\bibinfo{year}{2017}).

\bibitem[{\citenamefont{Ji et~al.}(2018)\citenamefont{Ji, Li, Ma, and
  Suzuki}}]{chueng_1}
\bibinfo{author}{\bibfnamefont{C.-R.} \bibnamefont{Ji}},
  \bibinfo{author}{\bibfnamefont{Z.}~\bibnamefont{Li}},
  \bibinfo{author}{\bibfnamefont{B.}~\bibnamefont{Ma}}, \bibnamefont{and}
  \bibinfo{author}{\bibfnamefont{A.~T.} \bibnamefont{Suzuki}},
  \bibinfo{journal}{Phys. Rev. D} \textbf{\bibinfo{volume}{98}},
  \bibinfo{pages}{036017} (\bibinfo{year}{2018}),
  \urlprefix\url{https://link.aps.org/doi/10.1103/PhysRevD.98.036017}.

\bibitem[{\citenamefont{Collins}(2018)}]{Collins:2018aqt}
\bibinfo{author}{\bibfnamefont{J.}~\bibnamefont{Collins}}
  (\bibinfo{year}{2018}), \eprint{1801.03960}.

\bibitem[{\citenamefont{Mannheim et~al.}(2021)\citenamefont{Mannheim, Lowdon,
  and Brodsky}}]{Mannheim:2020rod}
\bibinfo{author}{\bibfnamefont{P.~D.} \bibnamefont{Mannheim}},
  \bibinfo{author}{\bibfnamefont{P.}~\bibnamefont{Lowdon}}, \bibnamefont{and}
  \bibinfo{author}{\bibfnamefont{S.~J.} \bibnamefont{Brodsky}},
  \bibinfo{journal}{Phys. Rept.} \textbf{\bibinfo{volume}{891}},
  \bibinfo{pages}{1} (\bibinfo{year}{2021}), \eprint{2005.00109}.

\bibitem[{\citenamefont{Ma and Ji}(2021)}]{chueng_2}
\bibinfo{author}{\bibfnamefont{B.}~\bibnamefont{Ma}} \bibnamefont{and}
  \bibinfo{author}{\bibfnamefont{C.-R.} \bibnamefont{Ji}},
  \bibinfo{journal}{Phys. Rev. D} \textbf{\bibinfo{volume}{104}},
  \bibinfo{pages}{036004} (\bibinfo{year}{2021}),
  \urlprefix\url{https://link.aps.org/doi/10.1103/PhysRevD.104.036004}.

\bibitem[{\citenamefont{Polyzou}(2021)}]{polyzou:2021}
\bibinfo{author}{\bibfnamefont{W.~N.} \bibnamefont{Polyzou}},
  \bibinfo{journal}{Phys. Rev.} \textbf{\bibinfo{volume}{D103}},
  \bibinfo{pages}{105017} (\bibinfo{year}{2021}).

\bibitem[{\citenamefont{Brodsky et~al.}(2022)\citenamefont{Brodsky, Deur, and
  Roberts}}]{brodsky_2022}
\bibinfo{author}{\bibfnamefont{S.~J.} \bibnamefont{Brodsky}},
  \bibinfo{author}{\bibfnamefont{A.}~\bibnamefont{Deur}}, \bibnamefont{and}
  \bibinfo{author}{\bibfnamefont{C.}~\bibnamefont{Roberts}},
  \bibinfo{journal}{Nature Rev. Phys.} \textbf{\bibinfo{volume}{4}},
  \bibinfo{pages}{489} (\bibinfo{year}{2022}).

\bibitem[{\citenamefont{Gelfand et~al.}(1964)\citenamefont{Gelfand, Shilov,
  Graev, Vilenkin, and Pyatetskii-Shapiro}}]{gelfand}
\bibinfo{author}{\bibfnamefont{I.~M.} \bibnamefont{Gelfand}},
  \bibinfo{author}{\bibfnamefont{G.~E.} \bibnamefont{Shilov}},
  \bibinfo{author}{\bibfnamefont{M.~I.} \bibnamefont{Graev}},
  \bibinfo{author}{\bibfnamefont{N.~Y.} \bibnamefont{Vilenkin}},
  \bibnamefont{and} \bibinfo{author}{\bibfnamefont{I.~I.}
  \bibnamefont{Pyatetskii-Shapiro}}, \emph{\bibinfo{title}{{Generalized
  functions}}}, AMS Chelsea Publishing (\bibinfo{publisher}{Academic Press},
  \bibinfo{address}{New York, NY}, \bibinfo{year}{1964}), \bibinfo{note}{trans.
  from the Russian, Moscow, 1958},
  \urlprefix\url{https://cds.cern.ch/record/105396}.

\bibitem[{\citenamefont{Streater and Wightman}(1980)}]{Wightman:1980}
\bibinfo{author}{\bibfnamefont{R.~F.} \bibnamefont{Streater}} \bibnamefont{and}
  \bibinfo{author}{\bibfnamefont{A.~S.} \bibnamefont{Wightman}},
  \emph{\bibinfo{title}{PCT, Spin and Statistics, and All That}}
  (\bibinfo{publisher}{Princeton Landmarks in Physics}, \bibinfo{year}{1980}).

\bibitem[{\citenamefont{Haag}(1958)}]{Haag:1958vt}
\bibinfo{author}{\bibfnamefont{R.}~\bibnamefont{Haag}}, \bibinfo{journal}{Phys.
  Rev.} \textbf{\bibinfo{volume}{112}}, \bibinfo{pages}{669}
  (\bibinfo{year}{1958}).

\bibitem[{\citenamefont{Brenig and Haag}(1959)}]{Brenig:1959}
\bibinfo{author}{\bibfnamefont{W.}~\bibnamefont{Brenig}} \bibnamefont{and}
  \bibinfo{author}{\bibfnamefont{R.}~\bibnamefont{Haag}},
  \bibinfo{journal}{Fort. der Physik} \textbf{\bibinfo{volume}{7}},
  \bibinfo{pages}{183} (\bibinfo{year}{1959}).

\bibitem[{\citenamefont{Ruelle}(1962)}]{Ruelle:1962}
\bibinfo{author}{\bibfnamefont{D.}~\bibnamefont{Ruelle}},
  \bibinfo{journal}{Helv. Phys. Acta.} \textbf{\bibinfo{volume}{35}},
  \bibinfo{pages}{147} (\bibinfo{year}{1962}).

\bibitem[{\citenamefont{Jost}(1965)}]{jost:1966}
\bibinfo{author}{\bibfnamefont{R.}~\bibnamefont{Jost}},
  \emph{\bibinfo{title}{The General Theory of Quantized Fields (Lectures in
  Applied Mathematics, Volume IV}} (\bibinfo{publisher}{American Mathematical
  Society, Providence, Rhode Island, 1965.}, \bibinfo{year}{1965}).

\bibitem[{\citenamefont{Strocchi}(2013)}]{strocchi}
\bibinfo{author}{\bibfnamefont{F.}~\bibnamefont{Strocchi}},
  \emph{\bibinfo{title}{An Introduction to Non-Perturbative Foundations of
  Quantum Field Theory}} (\bibinfo{publisher}{Oxford University Press},
  \bibinfo{year}{2013}).

\bibitem[{\citenamefont{Reed and Simon}(1979)}]{simon}
\bibinfo{author}{\bibfnamefont{M.}~\bibnamefont{Reed}} \bibnamefont{and}
  \bibinfo{author}{\bibfnamefont{B.}~\bibnamefont{Simon}},
  \emph{\bibinfo{title}{Methods of Modern mathematical Physics}}, vol.
  \bibinfo{volume}{III Scattering Theory} (\bibinfo{publisher}{Academic Press},
  \bibinfo{year}{1979}).

\bibitem[{\citenamefont{Glaser et~al.}(1957)\citenamefont{Glaser, Lehmann, and
  Zimmermann}}]{Glaser:1957}
\bibinfo{author}{\bibfnamefont{V.}~\bibnamefont{Glaser}},
  \bibinfo{author}{\bibfnamefont{H.}~\bibnamefont{Lehmann}}, \bibnamefont{and}
  \bibinfo{author}{\bibfnamefont{W.}~\bibnamefont{Zimmermann}},
  \bibinfo{journal}{Il Nuovo Cimento} \textbf{\bibinfo{volume}{6}},
  \bibinfo{pages}{1122} (\bibinfo{year}{1957}).

\bibitem[{\citenamefont{Greenberg}(1965)}]{Greenberg:1965}
\bibinfo{author}{\bibfnamefont{W.~O.} \bibnamefont{Greenberg}},
  \bibinfo{journal}{Phys. Rev.} \textbf{\bibinfo{volume}{139}},
  \bibinfo{pages}{B1038} (\bibinfo{year}{1965}).

\bibitem[{\citenamefont{Nishijima}(1958)}]{nishijima}
\bibinfo{author}{\bibfnamefont{K.}~\bibnamefont{Nishijima}},
  \bibinfo{journal}{Phys. Rev.} \textbf{\bibinfo{volume}{111}},
  \bibinfo{pages}{995} (\bibinfo{year}{1958}).

\bibitem[{\citenamefont{Greenberg}(1978)}]{Greenberg:1978}
\bibinfo{author}{\bibfnamefont{O.~W.} \bibnamefont{Greenberg}},
  \bibinfo{journal}{Phys. Rev. D} \textbf{\bibinfo{volume}{17}},
  \bibinfo{pages}{2576} (\bibinfo{year}{1978}).

\bibitem[{\citenamefont{Coleman}(2019)}]{coleman}
\bibinfo{author}{\bibfnamefont{S.}~\bibnamefont{Coleman}},
  \emph{\bibinfo{title}{{Quantum Field Theory, Lectures of Sidney Coleman}}}
  (\bibinfo{publisher}{World Scientific Pub.}, \bibinfo{address}{Hackensack,
  New Jersey}, \bibinfo{year}{2019}).

\bibitem[{\citenamefont{Wigner}(1939)}]{Wigner:1939cj}
\bibinfo{author}{\bibfnamefont{E.~P.} \bibnamefont{Wigner}},
  \bibinfo{journal}{Annals Math.} \textbf{\bibinfo{volume}{40}},
  \bibinfo{pages}{149} (\bibinfo{year}{1939}).

\bibitem[{\citenamefont{Reed and Simon}(1972)}]{Simon_I}
\bibinfo{author}{\bibfnamefont{M.}~\bibnamefont{Reed}} \bibnamefont{and}
  \bibinfo{author}{\bibfnamefont{B.}~\bibnamefont{Simon}},
  \emph{\bibinfo{title}{Methods of Modern mathematical Physics}}, vol.
  \bibinfo{volume}{I Functional Analysis} (\bibinfo{publisher}{Academic Press},
  \bibinfo{year}{1972}).

\bibitem[{\citenamefont{Dirac}(1949)}]{Dirac:1949cp}
\bibinfo{author}{\bibfnamefont{P.~A.~M.} \bibnamefont{Dirac}},
  \bibinfo{journal}{Rev. Mod. Phys.} \textbf{\bibinfo{volume}{21}},
  \bibinfo{pages}{392} (\bibinfo{year}{1949}).

\bibitem[{\citenamefont{Sokolov and Shatnyi}(1979)}]{Sokolov:1977im}
\bibinfo{author}{\bibfnamefont{S.~N.} \bibnamefont{Sokolov}} \bibnamefont{and}
  \bibinfo{author}{\bibfnamefont{A.~N.} \bibnamefont{Shatnyi}},
  \bibinfo{journal}{Theor. Math. Phys.} \textbf{\bibinfo{volume}{37}},
  \bibinfo{pages}{1029} (\bibinfo{year}{1979}).

\bibitem[{\citenamefont{Keister and Polyzou}(1997)}]{Keister:1996bd}
\bibinfo{author}{\bibfnamefont{B.~D.} \bibnamefont{Keister}} \bibnamefont{and}
  \bibinfo{author}{\bibfnamefont{W.~N.} \bibnamefont{Polyzou}},
  \bibinfo{journal}{J. Comput. Phys.} \textbf{\bibinfo{volume}{134}},
  \bibinfo{pages}{231} (\bibinfo{year}{1997}), \eprint{nucl-th/9611049}.

\bibitem[{\citenamefont{Bogoliubov and Shirkov}(1959)}]{bogoliubov:1959}
\bibinfo{author}{\bibfnamefont{N.~N.} \bibnamefont{Bogoliubov}}
  \bibnamefont{and} \bibinfo{author}{\bibfnamefont{D.~V.}
  \bibnamefont{Shirkov}}, \emph{\bibinfo{title}{Introduction to the theory of
  quntized fields}} (\bibinfo{publisher}{Wiley}, \bibinfo{year}{1959}).

\bibitem[{\citenamefont{Araki}(1999)}]{araki-bk}
\bibinfo{author}{\bibfnamefont{H.}~\bibnamefont{Araki}},
  \emph{\bibinfo{title}{Mathematical theory of quantum fields}}
  (\bibinfo{publisher}{Oxford University Press}, \bibinfo{year}{1999}).

\bibitem[{\citenamefont{Zimmermann}(1958)}]{Zimmermann:1958}
\bibinfo{author}{\bibfnamefont{W.}~\bibnamefont{Zimmermann}},
  \bibinfo{journal}{Il Nuove Cimento} \textbf{\bibinfo{volume}{X}},
  \bibinfo{pages}{597} (\bibinfo{year}{1958}).

\bibitem[{\citenamefont{Yang and Feldman}(1950)}]{yangfeldman}
\bibinfo{author}{\bibfnamefont{C.}~\bibnamefont{Yang}} \bibnamefont{and}
  \bibinfo{author}{\bibfnamefont{D.}~\bibnamefont{Feldman}},
  \bibinfo{journal}{Phys. Rev.} \textbf{\bibinfo{volume}{79}},
  \bibinfo{pages}{972} (\bibinfo{year}{1950}).

\bibitem[{\citenamefont{Veltman}(1994)}]{Veltman}
\bibinfo{author}{\bibfnamefont{M.}~\bibnamefont{Veltman}},
  \emph{\bibinfo{title}{{Diagrammatica}}} (\bibinfo{publisher}{Cambridge
  University Press}, \bibinfo{address}{Melbourne, Australia},
  \bibinfo{year}{1994}).

\end{thebibliography}
\end{document}